\documentclass{article}

\usepackage{graphicx}
\usepackage{color}

\makeatletter
\def\maxwidth{ %
  \ifdim\Gin@nat@width>\linewidth
    \linewidth
  \else
    \Gin@nat@width
  \fi
}
\makeatother

\definecolor{fgcolor}{rgb}{0.345, 0.345, 0.345}

\usepackage[utf8]{inputenc}
\usepackage{framed}
\makeatletter
 {\par\unskip\endMakeFramed%
 \at@end@of@kframe}
\makeatother

\definecolor{shadecolor}{rgb}{.97, .97, .97}
\definecolor{messagecolor}{rgb}{0, 0, 0}
\definecolor{warningcolor}{rgb}{1, 0, 1}
\definecolor{errorcolor}{rgb}{1, 0, 0}

\usepackage{alltt}
\usepackage[intoc]{nomencl}

\textfloatsep = 0.4in \addtocontents{toc}{\vspace{0.4in} \hfill
Page\endgraf} \addtocontents{lof}{\vspace{0.2in} \hspace{0.13in} \
Figure\hfill Page\endgraf} \addtocontents{lot}{\vspace{0.2in}
\hspace{0.13in} \ Table\hfill Page\endgraf}
\usepackage{float}
\usepackage{textcomp,bm}
\usepackage{array}
\usepackage{listings}
\usepackage{pdfpages}
\usepackage{setspace}
\usepackage{mathptmx}
\usepackage{capt-of}
\usepackage[table, svgnames]{xcolor}
\usepackage{colortbl}
\usepackage{graphicx}
\usepackage{amssymb, amsmath}
\usepackage{subfig}
\usepackage{wrapfig}
\usepackage{epsfig}
\usepackage{times}
\usepackage{authblk}
\usepackage{float}
\usepackage{epstopdf}
\usepackage{rotating}
\usepackage{bm}
\usepackage{comment}
\usepackage{lipsum}
\usepackage{rotating}
\usepackage{makeidx}
\usepackage{url}
\usepackage{multirow}
\usepackage{pdfpages,caption}
\usepackage{booktabs}
\usepackage[subfigure, titles]{tocloft}
\usepackage{caption}
\usepackage{acronym}
\usepackage{datetime}
\usepackage{pdfpages}
\usepackage{lmodern}
\usepackage[english]{babel}
\usepackage[autostyle]{csquotes}
\usepackage[linktocpage=true]{hyperref}
\usepackage{amsmath}
\usepackage[font=small,skip=0pt]{caption}
 \usepackage[margin=1in]{geometry}
\usepackage{algorithm}
\usepackage{lscape}
\usepackage{mdframed}

\makenomenclature

\usepackage{makecell}
\usepackage{titletoc}
\usepackage{appendix}

\title{Estimating maternal mortality using data from national civil registration vital statistics systems: A Bayesian hierarchical bivariate random walk model to estimate sensitivity and specificity of reporting}

\author[1]{Emily Peterson}
\author[2]{Doris Chou}
\author[2]{Ann-Beth Moller}
\author[3]{Alison Gemmill}
\author[2]{Lale Say}
\author[1]{Leontine Alkema}

\affil[]{1. Department of Biostatistics and Epidemiology, University of Massachusetts Amherst, USA. 2. United Nations Maternal Mortality Estimation Intergency Group, World Health Health Organization, Geneva, Switzerland. 3. Johns Hopkins University, USA.}

\usepackage[nottoc]{tocbibind}
\IfFileExists{upquote.sty}{\usepackage{upquote}}{}

\begin{document}
\pagenumbering{alph}


\pagenumbering{roman} \setcounter{page}{1}
\captionsetup{font=small,skip=0pt}
\maketitle
\singlespacing
\begin{abstract}
\noindent
Civil registration vital statistics (CRVS) data are used to produce national estimates of maternal mortality, but are often subject to substantial reporting errors due to misclassification of maternal deaths. The accuracy of CRVS systems can be assessed by comparing CRVS-based counts of maternal and non-maternal deaths to those obtained from specialized studies, which are rigorous assessments of maternal mortality for a given country-period. We developed a Bayesian bivariate random walk model to estimate sensitivity and specificity of the reporting on maternal mortality in CRVS data, and associated CRVS adjustment factors. The model was fitted to a global data set of CRVS and specialized study data. Validation exercises suggest that the model performs well in terms of predicting CRVS-based proportions of maternal deaths for country-periods without specialized studies. The new model is used by the UN Maternal Mortality Inter-Agency Group to account for misclassification errors when estimating maternal mortality using CRVS data.
\end{abstract}
\bigskip

\noindent
\small{\textbf{Acknowledgments:} The authors are very grateful to all members of the United Nations Maternal Mortality Estimation Inter-Agency Group for the support of this work and technical advisors Saiffuddin Ahmed, Peter Byass, Thomas Pullum, Tim Colbourn, Jeff Eaton, Marie Klingberg Alvin, Helena Nordenstedt, Laina Mercer, and Jon Wakefield.  We are also grateful to the PAHO regional office collaborators Antonio Sanhueza, Patricia Lorena Ruiz Luna and Adrienne Lavita Cox. Additional thanks to Michael Lavine, Laura Balzer, and Ken Kleinman for specific comments and suggestions related to the CRVS adjustment model. We thank the numerous participants and staff involved in the collection and publication of the data that we analyzed. We thank Maria Barriex for help with the systematic review.\\

\noindent
\small{\textbf{Funding:} Financial support was provided by WHO, through the Department of Reproductive Health and Research and HRP (the UNDP-UNFPA-UNICEF-WHO-World Bank Special Programme of Research, Development and Research Training in Human Reproduction), USAID, and University of Massachusetts  Amherst. \\
\noindent
\small{\textbf{License: CC BY IGO.}\\
The views expressed in this paper are those of the authors and do not necessarily reflect the views of the WHO, UNICEF, UNFPA, the World Bank, or the United National Population Division.}

\clearpage


\tableofcontents

\begingroup
\setlength{\parskip}{1\baselineskip}
\printnomenclature
\newpage
\endgroup

\normalsize

\pagenumbering{arabic}
\setcounter{page}{1}
\singlespacing

 \section{Introduction}
 \noindent
The United Nations Maternal Mortality Estimation Inter-agency group (UN MMEIG) is responsible for publishing internationally comparable estimates of maternal mortality for UN Development Goal reporting. The 2019 UN MMEIG estimates were constructed using a Bayesian hierarchical time series regression model, referred to as BMat (UN MMEIG 2019). This model uses an input database which is based upon nationally representative data available from Civil Registration Vital Statistics (CRVS), population-based surveys such as DHS and MICS, censuses, and specialized surveillance. A more general explanation of these data sources and their limitations is included in the UN MMEIG 2019 report (UN MMEIG 2019). One main concern for countries and the UN MMEIG has been the process to account for potential errors associated with assigning cause of death or under-registration of deaths within CRVS-derived data inputs. Therefore, the main refinement made to BMat since its first use in 2015 (Alkema et al. 2015, UN MMEIG 2015, Alkema et al. 2017) regards the estimation of maternal mortality using data from CRVS systems. This paper describes the update in detail and provides technical details. A more general explanation is included in the UN MMEIG 2019 report (UN MMEIG 2019).\\
\noindent

National vital registration systems record the number of deaths to women of reproductive ages, as well as the cause associated with each death using ICD coding. Based on the number of all-cause and maternal deaths, the proportion of deaths that are of a maternal cause, referred to as the proportion maternal (PM), can be constructed. Under ideal circumstances, when all deaths are captured and all causes are accurately classified, VR systems provide perfect information on the number of maternal deaths within the country. However, even if routine registration of deaths is in place, maternal deaths may be reported incorrectly if deaths are unregistered or misclassified, where misclassification of deaths refers to incorrect coding in vital registration systems, due either to error in the medical certification of cause of death or error in applying the correct ICD code.\\
\noindent

The accuracy of CRVS systems can be assessed by comparing CRVS-based observed PMs to those obtained from specialized studies, which are rigorous assessments of maternal mortality for a given country-period. Prior work comparing the ratio of study-based PMs to CRVS-based PMs, referred to as CRVS adjustment factors, found that these ratios are around 150\%, thus suggesting that PMs obtained from CRVS do not adequately capture all maternal deaths (Wilmoth et al., 2012). Based on these findings, the UN MMEIG has applied adjustments to CRVS data, to reduce bias in CRVS-based derived data in settings where CRVS systems are subject to error. Specifically, for countries with specialized studies that overlapped with CRVS data, adjustments were calculated directly from available data (i.e. the ratio of the study’s reported PM to CRVS-based PM) and kept constant in extrapolations. For countries without specialized studies data, a global adjustment of 1.5 was applied.\\
\noindent

In this round, we developed a model, referred to as the CRVS adjustment model or CRVS model, to estimate the effect of misclassification errors on the reported CRVS-based PM for country-periods without information from specialized studies. The next section introduces terminology and the framework used to describe errors in reporting of maternal mortality in CRVS systems. Section 3 provides information on the data available to inform estimation of the extent of incorrect reporting. Section~\ref{sec:methods} introduces the a Bayesian model to estimate the extend of misclassification in the reporting of maternal deaths in CRVS systems. The estimation is based on summarizing misclassification in terms of sensitivity and specificity, and modeling these two indicators for all country-years with CRVS data using a bivariate hierarchical random walk model. Finally, we present findings and the results of validation exercises.

 \section{Reporting errors in CRVS systems}\label{defn}
 The diagram in Figure \ref{fig:6box} illustrates the breakdown of total deaths to women of reproductive age by CRVS-reporting status (columns) and true maternal cause (rows). In a complete-CRVS setting, meaning that all deaths are registered, the number of missed deaths (3rd column) is equal to zero, such that reporting errors are solely due to misclassification of deaths. Inaccurate attribution of cause of death is either due to error in the medical certification of cause of death, and/or error in applying the correct code, which results in two misclassification biases regarding maternal deaths. Firstly, error occurs when a maternal death is misclassified as non-maternal, referred to as a false negative $(F^-)$ maternal death. Secondly, if a non-maternal death is misclassified as maternal, the death is labeled as a false positive maternal death $(F^+)$. Correctly classified maternal and non-maternal deaths are indicated by true positive $(T^+)$ and true negative $(T^-)$ maternal deaths, respectively. From the individual categories in Figure \ref{fig:6box}, cumulative totals are calculated summing across rows and columns, i.e. CRVS reported maternal deaths is the sum of $T^{(+)}$ and $F^{(+)}$, whereas, the true number of maternal deaths within the CRVS is the sum of $T^{(+)}$ and $F^{(-)}$. In incomplete CRVS systems, missed deaths include unregistered maternal deaths, referred to as $U^{(+)}$ deaths, and unregistered non-maternal deaths $U^{(-)}$.

 \begin{figure}[H]
 	\includegraphics[width=0.9\textwidth]{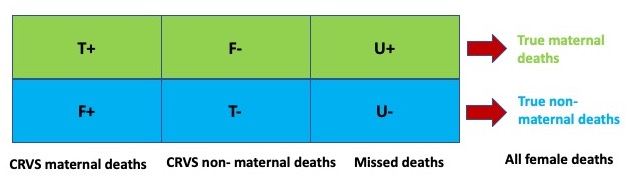}
 	\captionsetup[figure]{font=small,skip=0pt}
 	\caption{Diagram of breakdown of total deaths to women of reproductive age for a country-year, by CRVS-reporting status (columns) and true maternal cause (rows). $T^{(+)}$ and $F^{(-)}$ deaths refer to maternal deaths that are correctly registered as maternal deaths, and incorrectly registered as non-maternal deaths, respectively. Similarly, $F^{(+)}$ and $T^{(-)}$ maternal deaths refers to non-maternal deaths that are incorrectly registered as maternal deaths, and correctly registered as non-maternal deaths, respectively. $U^{(+)}$ refers to unregistered maternal deaths, and $U^{(-)}$ refers to unregistered non-maternal deaths. }
 	\label{fig:6box}
 \end{figure}

 \bigskip

\section{Data}\label{sec-data}
Information on CRVS misclassification errors and unregistered deaths was obtained from comparing information from specialized studies to CRVS reported deaths. This section discusses both types of data.

\subsection{CRVS data and completeness assessment}
The WHO Mortality Database maintains data from CRVS systems\footnote{WHO Mortality Database: https://www.who.int/healthinfo/mortality\_data/en/.}. Using this database, we obtained information on the number of maternal deaths reported in the CRVS and the number of deaths to women aged 15-49 reported in the CRVS (CRVS envelope).\\
\noindent
Completeness of the reporting of deaths into the CRVS system was assessed by comparing CRVS reported deaths to WHO estimates of deaths to women of reproductive age, obtained from life tables for WHO Member States\footnote{Life tables; Global Health Observatory (GHO) data. Geneva: World Health Organization; 2019 (https://www.who.int/gho/mortality\_burden\_disease/life\_tables/life\_tables/en/, accessed 18 June 2019).}. We first calculated the annual ratio of female deaths reported in the CRVS over deaths estimated by the WHO for all years with CRVS data, based on a moving window of 5-year periods (five-year periods were used to obtain less variable ratios for countries with smaller populations). If the ratios, more specifically, the upper bounds of 95\% confidence intervals when accounting for stochastic uncertainty in the ratio, are greater than 0.95 for all years with CRVS data, we assumed that the CRVS was complete in the country during the entire period. Otherwise, CRVS completeness was given by the ratio for each individual year (UN MMEIG 2019).

\subsection{Specialized studies}
A specialized study is defined as the assessment of maternal mortality for a country-period, either independent of CRVS reported data or based on the checking of CRVS reported deaths. These studies provided counts of the number of true maternal deaths (first row in Figure 1) or possibly individual categories, i.e. the number of false negative maternal deaths. We assumed that the study envelope was equal to the envelope reported by the CRVS system, unless specified otherwise in the study.
Specialized studies were obtained through (1) a literature review, (2) the UN MMEIG 2015 maternal mortality data base (UN MMEIG 2015), and (3) information provided by countries based on a follow-up survey, sent to countries in response to discussions with the Pan-American Health Organization (PAHO), and during country consultation. Detailed information on the compilation of specialized studies data is given in Appendix Section \ref{sec:studiesdatabase}.\\
\noindent

\subsection{Data availability}

A total of 50 study documents contributed data to inform the CRVS adjustment model, referring to 33 unique countries and 221 unique country-periods (observations). The majority of included study documents were obtained through the systematic search $(n = 22)$. In addition, 18 study documents were obtained from the UN MMEIG 2015 database (UN MMEIG 2015). Additional information from follow-up surveys and communication with countries during country consultation yielded 10 additional study documents (Appendix Section \ref{sec:studiesdatabase}).

Reported information varied greatly across observations. While some studies reported a detailed breakdown of false positive and/or false negative maternal deaths, the majority of studies reported only the confirmed total number of maternal deaths for a given country-period, see Table~\ref{table: tabdata}. The majority of studies reported on the true number of maternal deaths within the CRVS (184 observations, 30 countries). Information on both false negative and false positive breakdowns was available for 18 observations (4 countries). Most studies with breakdown information solely reported on false negative breakdowns, 38 observations from 4 countries. Data regarding the relative difference between the proportion of maternal deaths among CRVS-reported deaths and the proportion of maternal deaths among unregistered deaths was very limited: only 13 observations reported information that included $U^+$. 
\begin{table}[H]
	\begin{tabular}{|p{8cm}||p{3cm}|p{3cm}|}
		\hline
		Reported counts & \# of observations & \# of countries \\
		\hline 	\hline
		True maternal in CRVS only &  162 &  27\\ \hline
	True maternal in CRVS and U+ & 2 & 1 \\  \hline
		F- and F+ 	and U+ & 10  & 2 \\ \hline
		F- and F+  only & 8   & 2  \\ \hline
		F- and U+ only  &1 & 1\\ \hline
		F- only & 38 & 4\\
		\hline 	\hline
		Total & 221 & 33 \\
		\hline
	\end{tabular}
	\caption{Overview of data available from specialized studies. }
	\label{table: tabdata}
\end{table}

\section{Methods}\label{sec:methods}

\subsection{Summary of modeling approach }\label{sec:summary}
Based on the 6-box model, refer to Figure \ref{fig:6box}, for each country $c$ and year $t$, we assumed a multinomial data generating distribution as follows:
\begin{align}
\bm{y}_{c,t}|y_{c,t}^{(tot)}, \bm{\rho}_{c,t} &\sim Multinom \left(y_{c,t}^{(tot)}, \bm{\rho}_{c,t}\right),\label{eq-multi}
\end{align}
where (leaving out subscripts $(c,t)$ for improve readability):
\begin{align}
\bm{y}& = \left(y^{(U-)}, y^{(U+)}, y^{(T-)}, y^{(T+)}, y^{(F-)}, y^{(F+)}\right),\nonumber \\
\bm{\rho} & = \left(\rho^{(U-)}, \rho^{(U+)}, \rho^{(T-)}, \rho^{(T+)}, \rho^{(F-)}, \rho^{(F+)}\right), \nonumber
\end{align}
with $y^{(b)}$ the number of deaths reported for category $b$ in $B = \{T^+, T^-, F^+, F^-, U^+, U^-\}$ and $ y^{(tot)} = \sum_{b \in B} y^{(b)}$. Similarly, $\rho^{(b)}$ denotes the probability of a death in category $b$ and $\sum_{b \in B} \rho^{(b)} = 1$. Lastly, observed proportions are denoted with $p^{(b)} = y^{(b)}/y^{(tot)}$.
\noindent
Focusing on deaths captured in the CRVS data only, hence categories  $B^{(CRVS)} = \{T^+, T^-, F^+, F^-\}$, we define the total number of deaths in the CRVS as $y^{(CRVS)} = \sum_{b \in B^{(CRVS)}}y^{(b)}$, CRVS-based probabilities $\gamma^{(b)} = \gamma^{(b)}/\sum_{b \in B^{(CRVS)}}\rho^{(b)}$, and CRVS-based proportions $q^{(b)} = p^{(b)}/\sum_{b \in B^{(CRVS)}}p^{(b)}$. The proportion of CRVS-based deaths that is reported as being maternal (the CRVS-based observed PM) is given by $q^{(matCRVS)} = \left(y^{(T+)} + y^{(F+)}\right)/y^{(CRVS)}$.

\noindent
The question of interest is how to estimate the true probability of a maternal death, $\rho^{(truemat)} = \rho^{(T+)} + \rho^{(F-)} + \rho^{(U+)}$, based on the CRVS-reported maternal deaths $y^{(matCRVS)}$, total CRVS-reported deaths $y^{(CRVS)}$, and total deaths $y^{(tot)}$. Based on Eq.~\ref{eq-multi} we find
\begin{align}\label{eq:matCRVS}
y^{(matCRVS)}|y^{(CRVS)}, \gamma^{(matCRVS)} &\sim Bin \left(y^{(CRVS)}, \gamma^{(matCRVS)}\right),\\
\gamma^{(matCRVS)} &=  \left(\rho^{(T+)}+ \rho^{(F+)}\right)/\rho^{(CRVS)}, \nonumber
\end{align}
\noindent where $\gamma^{(matCRVS)}$ refers to the probability of reporting a death as being maternal in CRVS. For country-years with complete CRVS, this probability
$\gamma^{(matCRVS)}$ can be expressed as a function of  the true probability $\rho^{(truemat)}$, and misclassification parameters sensitivity $\lambda^{(+)}$ and specificity $\lambda^{(-)} $ as follows:
\begin{equation}\label{eq:truemat}
\gamma^{(matCRVS)}  = \lambda^{(+)}\rho^{(truemat)} + \left(1-\lambda^{(-)}\right)\left(1-\rho^{(truemat)}\right)
\end{equation}
\noindent
with sensitivity $\lambda^{(+)} = \frac{\gamma^{(T+)}}{\gamma^{(T+)} + \gamma^{(F-)}}$, the probability of correctly identifying a maternal death reported in the CRVS as such, and specificity $\lambda^{(-)} = \frac{\gamma^{(T-)}}{\gamma^{(T-)} + \gamma^{(F+)}}$, the probability of correctly identifying a non-maternal death  reported in the CRVS as such.
\bigskip

\noindent
For countries with incomplete CRVS systems, we define  $\omega^{(truematUNREG)}$ to be the probability of a maternal death among unregistered deaths, and $\gamma^{(truematCRVS)} = \gamma^{(T+)} + \gamma^{(F-)}$ to be the probability of a maternal death among CRVS-registered deaths. For these countries, Eq. \ref{eq:matCRVS} still holds true but the relation between $\gamma^{(matCRVS)}$ and $\rho^{(truemat)}$ changes if $\omega^{(truematUNREG)}$ differs from $\gamma^{(truematCRVS)}$. In such settings, the relation between $\gamma^{(matCRVS)}$ and $\rho^{(truemat)}$ can be written as follows:
\begin{equation}\label{eq-kappa}
\gamma^{(matCRVS)}=\frac{\lambda^{(+)} \cdot \rho^{(truemat)}}{\rho^{(CRVS)} + \left(1-\rho^{(CRVS)}\right) \cdot \kappa} + (1-\lambda^{(-)}) \cdot \left(1-\frac{\rho^{(truemat)}}{\rho^{(CRVS)} + (1-\rho^{(CRVS)})\cdot \kappa}\right),
\end{equation}
where $\kappa$ refers to the ratio of probabilities of a maternal death outside versus inside the CRVS:
$$\kappa = \frac{\omega^{(truematUNREG)}}{\gamma^{(truematCRVS)}}.$$

\noindent
We aimed to estimate sensitivity, specificity, and $\kappa$ (or a related parameter to summarize the relative difference between the probability of a maternal death outside versus inside the CRVS) for all country-years with CRVS data, such that CRVS data can be used to inform the estimation of maternal mortality among all deaths while accounting for CRVS misclassification errors and underregistration. However, given that data on the relative difference in maternal risk among CRVS-registered and unregistered deaths was so limited (see Table 1), we were unable to estimate this relative difference. Instead, we focused on the estimation of sensitivity and specificity using CRVS-based data only (221 observations, see Table 1). We developed a bivariate hierarchical random walk model for estimating sensitivity and specificity for all country-years, as explained in Section \ref{sec:proc}. We used all available CRVS-based data for model fitting, including data on the total number of maternal deaths only, as explained in Section~\ref{sec:modelfitting}.

\subsection{Bivariate hierarchical random walk model for sensitivity and specificity }\label{sec:proc}
\noindent
We developed a bivariate hierarchical random walk model to estimate sensitivity $\lambda_{c,t}^{(+)}$ and specificity $\lambda_{c,t}^{(-)}$ for all countries $c$ with CRVS data for some year(s) $t$. We constrained sensitivity (se) to be within 0.1 and 1, and specificity (sp) to be within 0.95 and 1 using transformations:
\begin{align*}
\eta^{(+)}_{c,t} &= log\left( \frac{\lambda_{c,t}^{(+)} - 0.1}{1-\lambda_{c,t}^{(+)}}\right),\\
\eta^{(-)}_{c,t} &= log\left( \frac{\lambda_{c,t}^{(-)} - 0.95}{1-\lambda_{c,t}^{(-)}}\right).\\
\end{align*}

\noindent
Sensitivity and specificity (after transformation) were modeled using bivariate distributions to account for possible correlation between the two misclassification parameters. Accounting for this correlation is important for estimating misclassification parameters, i.e. see Chu et al. 2006.
\noindent
The model set-up used is a hierarchical random walk process. In reference year $t_c$, here chosen as the midyear of the country-specific observation period, we assume a hierarchical distribution for transformed sensitivity and specificity:

\begin{align}
\begin{pmatrix}
\eta^{(+)}_{c,t_c}\\
\eta^{(-)}_{c,t_c} \end{pmatrix} & \sim N_2 \left(
\begin{bmatrix} \eta^{(+)}_{world} \\
\eta^{(-)}_{world} \end{bmatrix},
\begin{bmatrix} \sigma_{}^{(+)^{2}} &\phi\cdot \sigma_{}^{(+)}\cdot \sigma_{}^{(-)} \\
\phi\cdot \sigma_{}^{(+)}\cdot \sigma_{}^{(-)} & \sigma_{}^{(-)^{2}} \end{bmatrix}
\right) \label{eq-bhm}
\end{align}
\bigskip

\noindent
For years prior to the country-specific reference year, i.e. $t<t_c$:

\begin{align}
\begin{pmatrix}
\eta^{(+)}_{c,t}\\
\eta^{(-)}_{c,t} \end{pmatrix} & \sim N_2 \left(
\begin{bmatrix} \eta^{(+)}_{c,t+1} \\
\eta^{(-)}_{c,t+1} \end{bmatrix},
\begin{bmatrix} \delta^{(+)^{2}} &\phi\cdot \delta^{(+)}\cdot \delta^{(-)} \\
\phi\cdot \delta^{(+)}\cdot \delta^{(-)} & \delta^{(-)^{2}} \end{bmatrix}
\right). \label{eq-walk1}
\end{align}

\bigskip

\noindent
For years after the country-specific reference year, i.e. $t>t_c$:
\begin{align*}
\begin{pmatrix}
\eta^{(+)}_{c,t}\\
\eta^{(-)}_{c,t} \end{pmatrix} & \sim N_2 \left(
\begin{bmatrix} \eta^{(+)}_{c,t-1} \\
\eta^{(-)}_{c,t-1} \end{bmatrix},
\begin{bmatrix} \delta^{(+)^{2}} &\phi\cdot \delta^{(+)}\cdot \delta^{(-)} \\
\phi\cdot \delta^{(+)}\cdot \delta^{(-)} & \delta^{(-)^{2}} \end{bmatrix}
\right).\label{eq-walk2}
\end{align*}

\bigskip

\noindent
The following prior distributions were assigned to the global mean parameters:
 \begin{align*}
\lambda^{(+)}_{world} &\sim Unif(0.1, 1),\\
\lambda^{(-)}_{world} & \sim Unif(0.995, 1),\\
\eta^{(+)}_{world} &= log\left(\left(\lambda^{(+)}_{world} - 0.1)/(1 - \lambda^{(+)}_{world} \right)\right),\\
\eta^{(-)}_{world} &= log \left(\left(\lambda^{(-)}_{world} - 0.1)/(1 - \lambda^{(-)}_{world} \right)\right),
\end{align*}
\noindent
Prior distributions for the correlation and standard deviations of the random walk were as follows:
  \begin{align}
\rho^{cor} &\sim Unif(-0.95, 0.95),\\
\sigma^{()} & \sim N_{T(0,\infty)}(0,1),\\
\delta^{()} & \sim N_{T(0,\infty)}(0,1),
\end{align}
where $N_{T(0,\infty)}(0,1)$ denotes a half-normal distribution (a truncated normal distribution with lower bound at 0).
\noindent

We explored the use of indicators gross domestic product (GDP), the general fertility rate (GFR), the proportion of ill-defined causes, CRVS completeness, and ICD coding (ICD10 or earlier) as possible covariates to inform estimates of sensitivity and specificity. However, exploratory analyses suggested no substantially meaningful relations
and were excluded from the final model. Illustrative plots are included in Appendix Section \ref{sec:covariate}.

\bigskip

\subsection{Model fitting }\label{sec:modelfitting}
\noindent

Our goal is to estimate sensitivity and specificity using data from all country-years with CRVS-based specialized study data. Based on the assumption of a multinomial data generating process from Eq.\ref{eq-multi}, we assumed the following data generating process for study counts from the $i$th study in country $c[i]$ in reference year $t[i]$:

\begin{align}
\bm{z}_{i}|z^{(CRVS)}_{i}, \bm{\gamma}_{c[i],t[i]} &\sim Multinom\left(z^{(CRVS)}_{i}, \bm{\gamma}_{c[i], t[i]}\right),\label{eq-studydm}
\end{align}
with study counts $\bm{z}_{i}= \left(z^{(T-)}_{i}, z^{(T+)}_{i}, z^{(F-)}_{i}, z^{(F+)}_{i}\right)$, $z^{(CRVS)}_{i} = \sum_{b \in B^{(CRVS)}} z_{i}^{(b)}$, and unknown probability vector $\bm{\gamma}_{c,t} = \left(\gamma^{(T-)}_{c,t}, \gamma^{(T+)}_{c,t}, \gamma^{(F-)}_{c,t}, \gamma^{(F+)}_{c,t}\right)$. For studies that refer to one calendar year, the study counts corresponds to the counts for that specific year, $z^{(b)}_{i} = y_{c[i],t[i]}^{(b)}$, while for studies that refer to multiple years, study counts are aggregates over the observation period, i.e., $z^{(b)}_{i} = \sum_{t=t1[i]}^{t2[i]} y_{c,t}^{(b)}$ where $t1[i]$ and $t2[i]$ refer to the start and end years of the $i$th study, respectively.

The 4 CRVS-based probabilities $\gamma^{(b)}_{c,t}$ can be written in terms of the two misclassification parameters $\lambda_{c,t}^{(+)}$ and $\lambda_{c,t}^{(-)}$, and the true CRVS-based probability of a maternal death as follows:
\begin{align*}
\gamma_{c,t}^{(T+)} &= \lambda_{c,t}^{(+)} \cdot \gamma_{c,t}^{(truemat)}, \\
\gamma_{c,t}^{(F-)} &= \gamma_{c,t}^{(truemat)} - \rho_{c,t}^{(T+)},\\
\gamma_{c,t}^{(T-)} &= \lambda_{c,t}^{(-)} \cdot \left(1-\gamma_{c,t}^{(truemat)}\right),\\
\gamma_{c,t}^{(F+)} &=  \left(1-\gamma_{c,t}^{(truemat)}\right)- \rho_{c,t}^{(T-)}.
\end{align*}
\noindent
Country-year model parameters are defined through the bivariate hierarchical random model on $\lambda_{c,t}^{(+)}$ and $\lambda_{c,t}^{(-)}$, and vague independent priors on $\gamma_{c,t}^{(truemat)}$:
$$\gamma_{c,t}^{(truemat)} \sim U(0, 1).$$

For studies that report on a specific set of non-overlapping categories, i.e. the number of false positive maternal deaths and/or the number of true positive maternal deaths, the corresponding likelihood function was obtained directly using the multinomial data generating process in Eq.~\ref{eq-studydm}.

\noindent
However, the majority of studies only reported information on the number of true maternal deaths within the CRVS (see table \ref{table: tabdata}). For each study that reported true maternal deaths within the CRVS, the study reported count of maternal deaths, $z_i^{(truematCRVS)}= z_i^{(T+)} + z_i^{(F-)}$, overlaps with the CRVS-reported maternal deaths for the corresponding country-period, $z_i^{(matCRVS)} = \sum_{t=t1[i]}^{t2[i]} y_{c[i],t[i]}^{(T+)} + y_{c[i],t[i]}^{(F+)}$. For each study period with information on overlapping categories, we obtained the exact likelihood function for the available death counts by summing over multinomial densities evaluated at each unique combination $\tilde{\bm{z}}_i = \left(\tilde{z}^{(T-)}_{i}, \tilde{z}^{(T+)}_{i}, \tilde{z}^{(F-)}_{i}, \tilde{z}^{(F+)}_{i}\right)$ that satisfied the observed set of counts. Specifically, for studies with information on the true number of maternal deaths $\left(z_i^{(truematCRVS)}\right)$, and the number of maternal deaths observed in the CRVS $\left(z_i^{(matCRVS)}\right)$, the likelihood function  $f_i = f\left(z_{i}^{(matCRVS)}, z_{i}^{(truematCRVS)}|z_{i}^{(CRVS)}, \bm{\gamma}_{c[i], t[i]}\right)$ is written as follows
\begin{align*}
f_i = \sum_{\tilde{z}_{i}^{(T+)}=0}^{z_{i}^{(matCRVS)}} p_z\left(\tilde{\bm{z}}|z_{i}^{(CRVS)}, \bm{\gamma}_{c[i], t[i]}\right)\cdot 1\left(\tilde{z}_{i}^{(T+)}+\tilde{z}_{i}^{(F-)} = z_{i}^{(truematCRVS)}\right)\cdot k_i
\end{align*}
where $p_z\left(\tilde{\bm{z}}|z_{i}^{(CRVS)}, \bm{\gamma}_{c[i], t[i]}\right)$ refers to the multinomial density function for the 4 CRVS-based categories from Eq.~\ref{eq-studydm}. Additionally, to improve computational efficiency and remove combinations that result in values of sensitivity and specificity with negligible probabilities, we added additional constraints to possible combinations of $\tilde{\bm{z}}_i$, reflected in $k_i$ with
$$k_i = 1\left(\tilde{z}_i^{(T+)} \geq Bin_{2.5\%}\left(z_i^{(truematCRVS)}, 0.1\right)\right) \\
\cdot  1\left(\tilde{z}_i^{(T-)} \geq Bin_{2.5\%}\left(z_{i}^{(CRVS)} - z_i^{(truematCRVS)}, 0.97 \right)\right)$$
where $Bin_{2.5\%}(n, p)$ refers to the 2.5th percentiles of a Binomial distribution with sample size $n$ and probability $p$, 0.1 is a lower bound for sensitivity, and 0.97 is a lower bound for specificity.

\bigskip

\subsection{Computation}
\noindent
A Markov Chain Monte Carlo (MCMC) algorithm was employed to sample from the posterior distribution of the parameters with the use of the software \textit{JAGS} (Plummer 2003). Ten parallel chains were run with a total of 40,000 iterations in each chain. Of these, the first of 10,000 iterations in each chain were discarded as burn-in and every 20th iteration after was retained. The resulting chains contained 1,500 samples each, with a total of 15,000 posterior samples. Standard diagnostic checks (using trace plots and Gelman and Rubin diagnostics (Gelman and Rubin 1992)) were used to check convergence.
\bigskip

\subsection{CRVS adjustment factor} \label{sec-adj}
Based on estimates of sensitivity and specificity, for countries with complete CRVS systems, we defined the associated CRVS adjustment factor for country $c$ in year $t$ as follows:
\begin{align}\label{eq.crvs}
CRVSadj_{c,t} &= \frac{p_{c,t}^{truemat}}{\lambda_{c,t}^{(+)} \cdot p_{c,t}^{(truemat)} + \left(1-\lambda_{c,t}^{(-)}\right)\cdot\left(1-p_{c,t}^{(truemat)}\right)},
\end{align}
which varies with the true PM $p_{c,t}^{truemat}$. For country-years without specialized studies, CRVS-adjustment factors follow from estimates of sensitivity and specificity, and the true PM.

\subsection{Comparison to UN MMEIG 2015 approach}\label{sec:bmatCRVS}
In the UN MMEIG 2015 approach, CRVS adjustment factors were obtained for all country-years with CRVS data and used directly in model fitting (Alkema et al. 2017). For countries with specialized studies, the CRVS adjustment in the UN MMEIG 2015 approach was calculated for country-periods with studies by taking the ratio of the study-based observed proportion of maternal deaths to the observed CRVS-based proportion (Alkema et al. 2017). Linear interpolation was used to obtain adjustments in years in between observed adjustments. For forward extrapolation, the CRVS adjustment was kept constant at the level of the most recent observed CRVS adjustment. Backward extrapolations are explained below. The uncertainty of the adjustment was set equal to the variability associated with $g$, defined as follows:
$$\log(g)|G \sim N\left(\log(G), 0.25^2\right),$$
where $G$ refers to the point estimate of the adjustment factor.

For countries with CRVS data but no specialized studies, the UN MMEIG used a constant global adjustment factor of 1.5 for all country-years (Wilmoth et al. 2012, Alkema et al. 2017). For backward extrapolations in countries with studies, the CRVS adjustment was assumed to increase or decrease linearly to the same global adjustment factor of 1.5 in 5 years.

The approach to obtaining CRVS adjustment with the CRVS-model differs from the UN MMEIG 2015 approach; the CRVS adjustment factor is obtained from estimates of sensitivity and specificity, and varies with the true PM, see Section~\ref{sec-adj}.
We explain how sensitivity, specificity, and the corresponding adjustments were obtained for all country-periods with CRVS data for the UN MMEIG 2019 estimates in Appendix Section \ref{sec:bmat}.

\subsection{Model validation}\label{validmethods}\label{sec-val}
\noindent
Model performance was assessed through two out-of-sample validation exercises. In the first exercise, $20\%$ of the observations were left out at random to form a training data set. The process was repeated 20 times, i.e. 20 training sets were constructed with different samples left out in each set. In the second exercise, we left out the observation corresponding to the most recent study period in each country. The CRVS adjustment model was fitted to each training set, and we obtained posterior samples for sensitivity and specificity in the country-years with left-out specialized studies.

To validate model performance, we combined samples of sensitivity and specificity with information on study-based observed PMs to obtain samples of predicted CRVS-based PMs. We summarized the difference  in terms of error, i.e., the difference between the observed CRVS-based PM and its point estimate, and coverage of 80\% prediction intervals. The procedure is described in detail in  Box~\ref{fig:validmeth}.
\singlespacing
\begin{figure}[H]
\fbox{
\begin{minipage}{45em}
\textbf{Calculation of outcome measures in the validation exercise}
\begin{enumerate}
\item Fit the CRVS adjustment model to the training data and obtain posterior samples $se^{(s)}_{c,t}$ and $sp^{(s)}_{c,t}$ for posterior samples $s=1,2,\hdots, S$ for country-years with left-out data in the test set.
\item Sample the CRVS-based reported number of maternal deaths using samples for sensitivity and specificity:
         \[z_i^{(matCRVS)^{(s)}}= z_i^{(truematCRVS)} \cdot se^{(s)}_{c[i], t[i]} + \left(1-z_i^{(truematCRVS)}\right) \cdot \left(1-sp^{(s)}_{c[i], t[i]}\right)\]
\item Calculate the difference between observed and estimated CRVS-based PM:
   \[error_i^{matCRVS^{(s)}} = \left(z_i^{(matCRVS)} - z_i^{(matCRVS)^{(s)}}\right)/z_i^{(CRVS)}\]
The median of the sampled errors is reported.
\item  Calculate the proportion of CRVS-based PMs $z_i^{(matCRVS)}/z_i^{(CRVS)}$ above and below their respective 80\% prediction interval.
\end{enumerate}
\end{minipage}
}
\caption{Overview of calculation of errors and coverage of prediction intervals in out-of-sample validation exercises.}\label{fig:validmeth}
\end{figure}

\bigskip

\section{Results}\label{results}
\subsection{Validation Results}
The CRVS adjustment model performs well in out-of-sample validation exercises, see Table~\ref{tab:valid}. Median and relative errors are small in both exercises, and absolute errors are around 10\% in predicting the CRVS-based PM. The model is well calibrated, the coverage of the 80\% prediction intervals is around 80\%, with around 10\% falling below (above) the lower (upper) bounds.

\begin{table}[H]
\small
\centering
\begin{tabular}{|p{2.5cm} |p{2cm} |p{1.5cm}|p{1.4cm} p{1.2cm}|p{1.2cm} p{1.2cm}|p{1.2cm} p{1.2cm}|  }
\hline
\multicolumn{9}{|c|}{\textbf{{Error in CRVS-PM}}} \\
\hline
\textbf{Validation} & \textbf{Model} & \textbf{$\#$ left-out obs} & \multicolumn{2}{|c|}{\textbf{Median Errors} } & \multicolumn{2}{|c|}{\textbf{Relative Error $(\%)$}} &  \multicolumn{2}{|c|}{\textbf{outside $80\%$ PI}}  \\
& & & ME & MAE & MRE & MARE & $\%$ Below & $\%$ Above\\
  \hline
Leave-out $20\%$ at random  & CRVSadj & 43  & 0.000009 & 0.0006  &  0.5 &  9.9  & 0.11  & 0.11 \\
 \hline
Leave-out last observation  & CRVSadj & 20  & 0.000340 & 0.0006 &  2.0 &  10.8  & 0.10  & 0.10 \\
 \hline
\end{tabular}
\caption{Validation results. The outcome measures are: median error (ME), median absolute error (MAE), relative error (MRE), absolute relative error (MARE), as well as the $\%$ of left-out observations below and above their respective $80\%$ prediction intervals (PI) based on the training set.}
\label{tab:valid}
\end{table}

\subsection{Global findings}
\noindent
Table \ref{tab:globals} lists the posterior estimates of the hyperparameters of the CRVS adjustment model. In the reference year, sensitivity is estimated at 0.586 ($80\%$ credible interval (CI) given by $(0.511, 0.656)$) and specificity is 0.9993 (0.9990, 0.9996).
The correlation between sensitivity and specificity was not significantly different from 0 (-0.095 $[-0.362, 0.183])$. There is substantial uncertainty associated with sensitivity and specificity in the reference year.
\bigskip

\begin{table}[ht]
\centering
\small
\begin{tabular}{|r|c|c|c|}
  \hline
 & 10\% & 50\% & 90\% \\
  \hline
global sensitivity$\lambda_{world}^{(+)}$ & 0.511 & 0.586 & 0.656 \\
global specificity$\lambda_{world}^{(-)}$ & 0.9990 & 0.9993 & 0.9996 \\
correlation $\phi$& -0.362 & -0.095 & 0.183 \\
sd sensitivity in $t_c$ ${\sigma}_{tref}^{(+)}$ & 0.915 & 1.161 & 1.490 \\
sd specificity in $t_c$  ${\sigma}_{tref}^{(-)}$ & 0.871 & 1.293 & 1.842\\
sd sensitivity in RW ${\delta}^{(+)}$ & 0.161 & 0.201 & 0.255 \\
sd specificity in RW $\delta^{(-)}$ & 0.508 & 0.673 & 0.857 \\
   \hline
\end{tabular}
\caption{Posterior estimates of global parameters; median estimate (50\%) and lower (10\%) and upper (90\%) bounds of $80\%$ credible intervals.}
   \label{tab:globals}
\end{table}

\noindent

Figure~\ref{fig:globplot} shows the relationship between true PM and the estimated CRVS adjustment factors, for specific values of specificity to illustrate their effect on the CRVS adjustment factor. When specificity equals one, the CRVS adjustment factor equals one over sensitivity, hence lower sensitivity results in a higher adjustment; conversely higher sensitivity results in a lower adjustment. When  specificity is less than one, while keeping sensitivity fixed, the adjustment factor decreases with decreasing true PM. This effect is due to an increasing share of false positive maternal deaths among all deaths, and a decreasing share of false negative deaths, or, in other words, as the true PM decreases, the proportion of non-maternal deaths reported as maternal increases while the proportion of maternal deaths reported as non-maternal decreases. This relationship implies that keeping specificity and sensitivity constant in extrapolations will result in changing adjustment factors as the true PM changes. Specifically, the adjustment factor will decrease if the true PM decreases in forward projections. Similarly, when using a fixed value of sensitivity and specificity, the adjustment factor associated with these values will depend on the value of the true PM.

\begin{figure}[H]
\center
\includegraphics[scale=0.5]{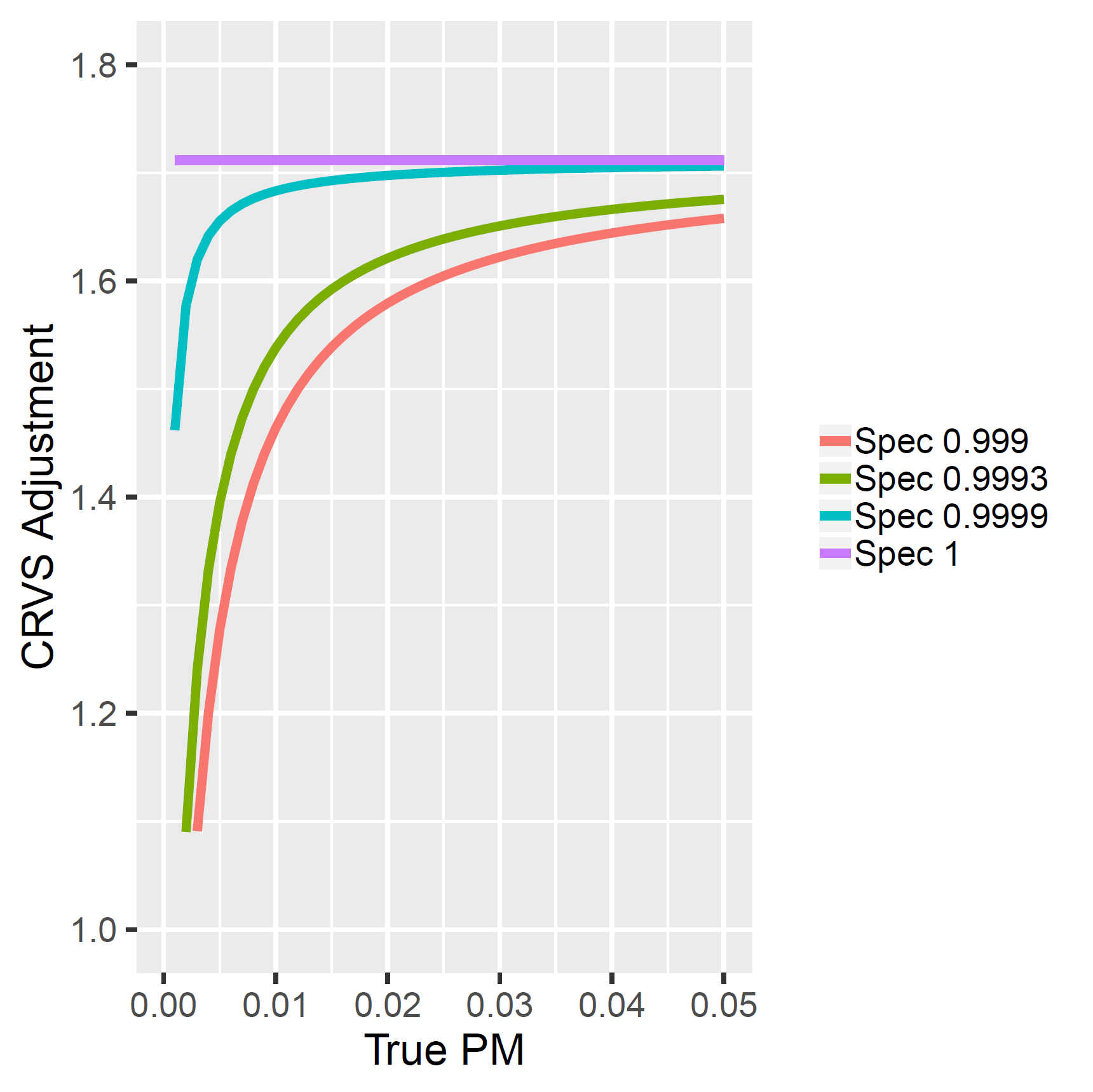}
\caption{CRVS adjustment for different values of specificity, calculated at different levels of true PM when sensitivity is fixed at the global estimate of 0.586.}
\label{fig:globplot}
\end{figure}

\subsection{Country estimates}\label{estresults}
\noindent
Sensitivity, specificity and CRVS adjustment estimates are shown for selected countries in Figure \ref{fig:subplot}. Posterior estimates (blue) are shown with observed data (red) during the estimation period.
The figure illustrates how uncertainty in estimates of sensitivity and specificity depends on (i) what information is available, (ii) the number of deaths in the country, and (iii) the observation years. In most countries, for example in Australia and the United Kingdom, the only available data is on true PM and CRVS-based PM across one or more periods. In these cases, sensitivity and specificity are unobserved, but are informed by observed data on true PM and CRVS-based PM.  This results in larger uncertainty bounds for sensitivity and specificity estimates as compared to the same setting but with available information on breakdowns. An example country with breakdown information is Brazil, where sensitivity and specificity are recorded for recent years, and estimated with less uncertainty. In addition to availability of data, the number of deaths in the country also determines the uncertainty in estimated sensitivity and specificity. For example, data in New Zealand is very uncertain due to the extremely small number of maternal deaths and total number of deaths to women of reproductive age. Lastly, uncertainty in sensitivity and specificity increases in years further away from years with data. This is illustrated in New Zealand, where data are available for recent years only; the uncertainty in sensitivity and specificity increases during periods without data.

\begin{sidewaysfigure}[ht]
    \includegraphics[scale=0.16]{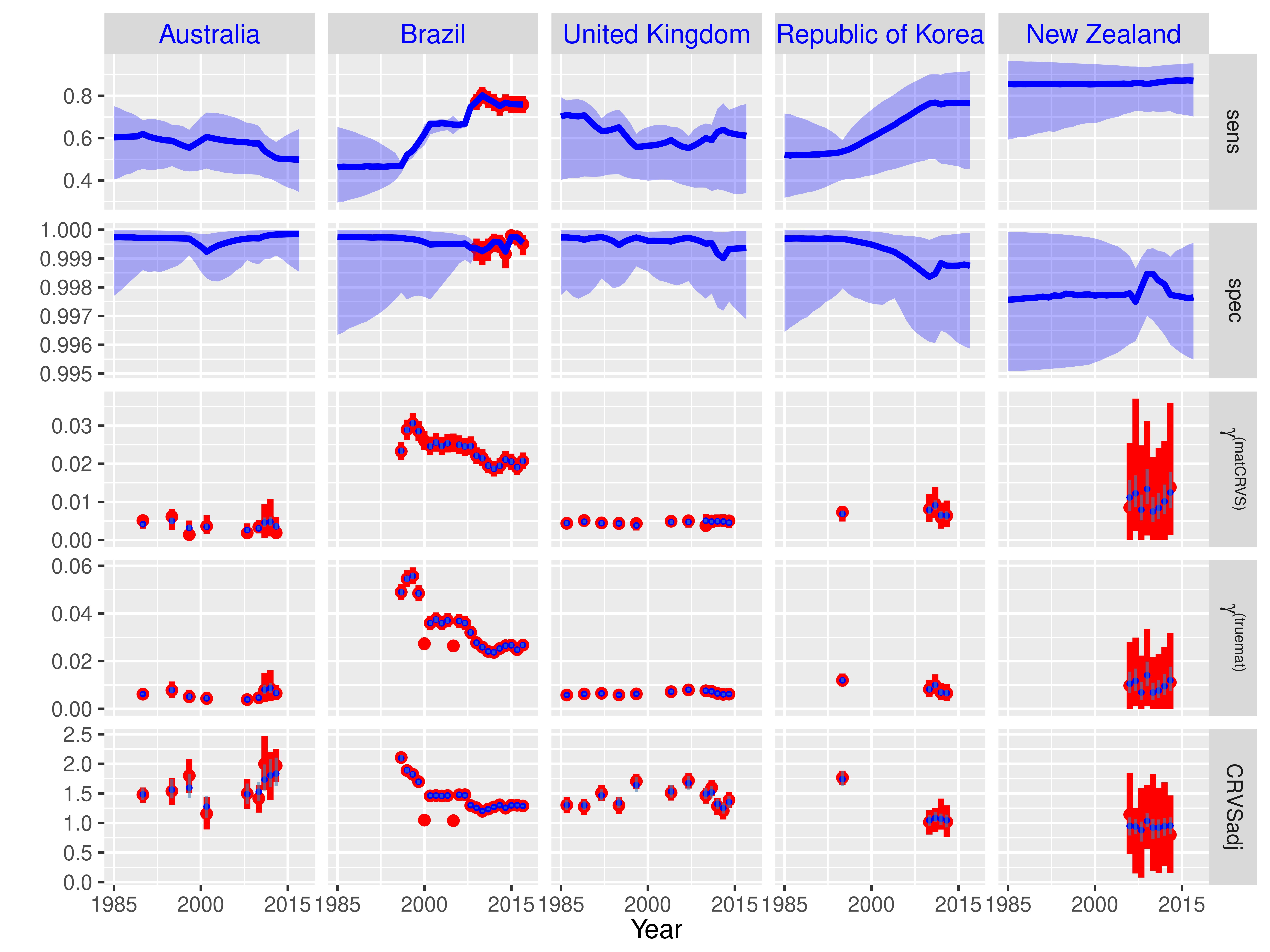}
   \caption{Illustration of CRVS adjustment model data and estimates for Australia, Brazil, the United Kingdom, Republic of Korea, and New Zealand. Parameters plotted consist of CRVS-based PM, true PM, sensitivity, specificity, and CRVS adjustment factors. The plots include: 1. observed data with associated observation-based 80\% confidence intervals (red), 2. posterior estimates with 80\% credible intervals (blue). }
    \label{fig:subplot}
\end{sidewaysfigure}

\clearpage


  \section{Summary}
In this paper, we presented a Bayesian hierarchical random walk model to assess maternal mortality misclassification errors in the CRVS with uncertainty. The model is based on the assessment of sensitivity and specificity of maternal mortality reporting, and captures differences therein between countries and within countries over time. Validation exercises suggest that the model performs well in terms of predicting CRVS-based PM for country-periods without specialized studies.\\

\noindent
The new model improves upon limitations of the 2015 UN MMEIG approach. In the UN MMEIG 2015 round of estimation, for countries with specialized studies that overlapped with CRVS data, adjustments were calculated directly from available data (i.e. the study’s reported PM to CRVS-based PM) and kept constant in extrapolations. The rationale for keeping adjustments constant in the 2015 approach for countries with studies was to implement ``no change in quality of reporting’’. However, when measuring quality of reporting in terms of sensitivity and specificity, the adjustment is not constant but varies with the true PM when keeping quality metrics constant, as illustrated in Figure~\ref{fig:globplot}. The CRVS model-based approach to obtaining adjustment factors improves upon this limitation of the UN MMEIG 2015 approach because its projections, which are based on constant sensitivity and specificity, are aligned with the assumption of constant quality of reporting. Similarly, for countries without specialized studies, the new approach uses global estimates of sensitivity and specificity to obtain adjustment factors, such that adjustment estimates are based on the same estimates of reporting accuracy, in contrast to using the same adjustment for all country-periods as in the UN MMEIG 2015 approach. Finally, uncertainty assessments differ between the old and new approach.  In the old approach, uncertainty in adjustments was assumed to be around 50\% for all country-periods. In the new approach, uncertainty in the adjustment factor follows from the uncertainty in the estimates for sensitivity and specificity and resulting adjustments are more certain in settings with recent information about quality of reporting.

\section{Appendix}
\subsection{Definitions}
\begin{table}[H]
\begin{tabular}{|p{3cm}|p{13cm}|}
\hline
\hline
\textbf{Term} & \textbf{Description}\\
\hline
\hline
Maternal death & The death of a woman whilst pregnant or within 42 days of termination of pregnancy, irrespective of the duration and site of the pregnancy, from any cause related to or aggravated by the pregnancy or its management but not from accidental or incidental causes define with the International Statistical Classification of Diseases and Related Health Problems 10th revision (ICD-10)\\
\hline
CRVS & Civil registration vital statistics, national death registration statistics\\ \hline
Specialized Study & (1) A study conducted precisely for the purpose of assessing the extent of misclassification within the CRVS and/or the extent of ``missingness" of maternal deaths, (2) A study conducted to independently assess cause of death classification among the true number of maternal deaths.\\
\hline
BMat & Bayesian maternal mortality estimation model, used by the UN MMEIG. BMat 2019 refers to the model used in the 2019 estimation round.\\
\hline
Sensitivity & (1) True positive rate, (2) Proportion of correctly classified maternal deaths to the true number of maternal deaths within CRVS systems.\\
\hline
Specificity & (1) True negative rate, (2) Proportion of correctly classified non-maternal deaths to the true number of non-maternal deaths within CRVS systems.\\
\hline
True positive maternal death & A maternal death correctly classified as maternal within CRVS.\\
\hline
True negative maternal death & A non-maternal death correctly classified as non-maternal within CRVS.\\
\hline
False positive maternal death & A non-maternal death misclassified as maternal within CRVS.\\
\hline
False negative maternal death & A maternal death misclassified as non-maternal within CRVS.\\
\hline
Missed/unregistered maternal death & A maternal death unregistered (missed) within CRVS, and therefore, unreported.\\
\hline
PM & The proportion of maternal deaths out of the total deaths to women of reproductive age (15-49).\\
\hline
CRVS-based PM & The proportion of CRVS reported maternal deaths out of the total deaths to women of reproductive age within CRVS.\\
\hline
CRVS adjustment & Relative adjustment needed to CRVS-based PM to obtain true PM.\\
\hline
\end{tabular}
\label{tab:defn}
\end{table}
\clearpage

\subsection{Compilation of specialized studies data}\label{sec:studiesdatabase}
\subsubsection{Summary of systematic review process}
\noindent
The objective of the review was to assess the level of misclassification reported by national official agencies for all WHO Member States. In other words, what is the level of incorrect reporting of maternal deaths in national official CRVS reporting, e.g. what is the difference between official reported number of maternal deaths versus the number of maternal deaths identified through special maternal mortality studies, confidential enquiries and surveillance systems etc. And to what extent is the incorrect reporting of maternal death due to misclassification versus missed or unregistered maternal deaths?

This review identified studies that fulfilled inclusion criteria as follows:
\begin{table}[H]
	\begin{tabular}{|p{3cm}|p{13cm}|}
		\hline
		\hline
		& \textbf{Inclusion Criteria}\\
		\hline
		\hline
		Population & Women of reproductive age (15-49 years) who died during pregnancy or up to one year after termination of pregnancy, irrespective of duration and the site of the pregnancy, from any cause.\\
		\hline
		Concept & Assessment of misclassification of maternal deaths by CRVS systems.\\
		\hline
		Study design & Cross-sectional study and retrospective cohort\\
		\hline
		Context & All WHO Member States reporting CRVS data\\
		\hline
	\end{tabular}
	\label{tab:criteria}
\end{table}
\noindent

In addition, the following criteria has to be met for inclusion:
\begin{enumerate}
	\item study is nationally representative;
	\item mid-years of reported data are after 1990;
	\item there is a matched comparison of CRVS data available in the study or in the WHO Mortality Database.
\end{enumerate}

\bigskip

\paragraph{Search Strategy}
\noindent
The search strategy was conducted for all relevant existing literature based on search terms relevant to the research questions restricted to the years 1990-2016, using the following online bibliographic databases: PubMed/MEDLINE, EMBASE, Global Index Medicus, EBSCO, Web of Science and Popline. The searches were conducted without any language restrictions. Search terms are included in Box at the end of this document. A hand search was also conducted on all WHO Member States Ministries of Health (MoH) websites to identify pertinent MoH maternal mortality and confidential inquiries reports.

\bigskip

\paragraph{Data Extraction}
\noindent
Data were extracted from full-text journal articles and reports which met the inclusion criteria. Data were extracted using a Microsoft Excel database. Information retrieved from the included studies included country, years assessed, study objectives, methodology /study design, number of maternal deaths, information on misclassification and incompleteness when available. Specifically, extraction focused on the assessment of the following:
\begin{enumerate}
\item The process by which the study retrieved and reviewed information on maternal deaths, including data source descriptions, definitions used by study, and whether the study reviewed all deaths to women of reproductive age or a description of the subset of deaths collected.
\item The number of maternal deaths, any information pertaining to misclassification of maternal cause of death by the CRVS system, any information regarding missed deaths by maternal cause.
\item Breakdown of maternal deaths by maternal cause of death was extracted if reported.
\end{enumerate}
\noindent

\subsubsection{Compilation of data}
The PRISMA diagram in Figure \ref{fig:prisma} provides information on the number of study documents and associated study observations both identified and included by (1) systematic review, (2) WHO maternal mortality database, and (3) information obtained from follow-up surveys and country consultation. Lastly, it reports the number of studies excluded and reason for exclusion at each stage of the screening process.  Studies were excluded in 3 subsequent steps. Firstly, studies were excluded if they reported information that could not be used, i.e. if no information on maternal death counts in the CRVS or associated envelopes could be obtained
(non-usable data). Secondly, a study was excluded if it was not nationally representative. Lastly, a study was excluded if an alternate study with more up-to-date or detailed information for the same country-period was available. The complete set of references of the included study documents is given in Box 1 at the end of this document.

\begin{figure}[H]
	\includegraphics[width=0.999\textwidth]{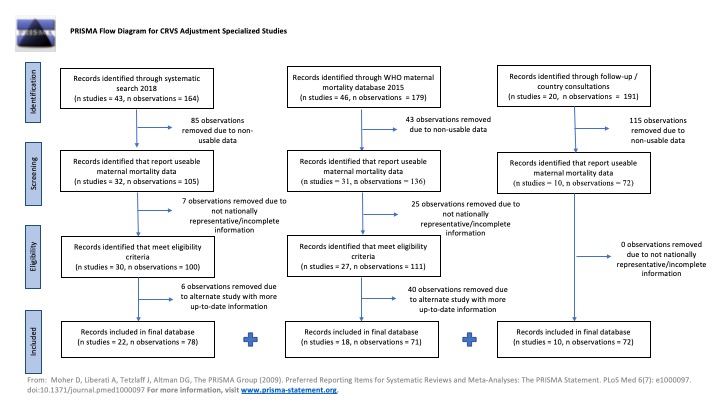}
	\caption{PRISMA flow diagram of data compilation of specialized studies for inclusion in the CRVS adjustment model. The numbers of studies mentioned refer to study documents. }
	\label{fig:prisma}
\end{figure}
\noindent

\clearpage

\subsection{Covariate plots}\label{sec:cov}\label{sec:covariate}
\begin{figure}[H]
	\center
	\includegraphics[]{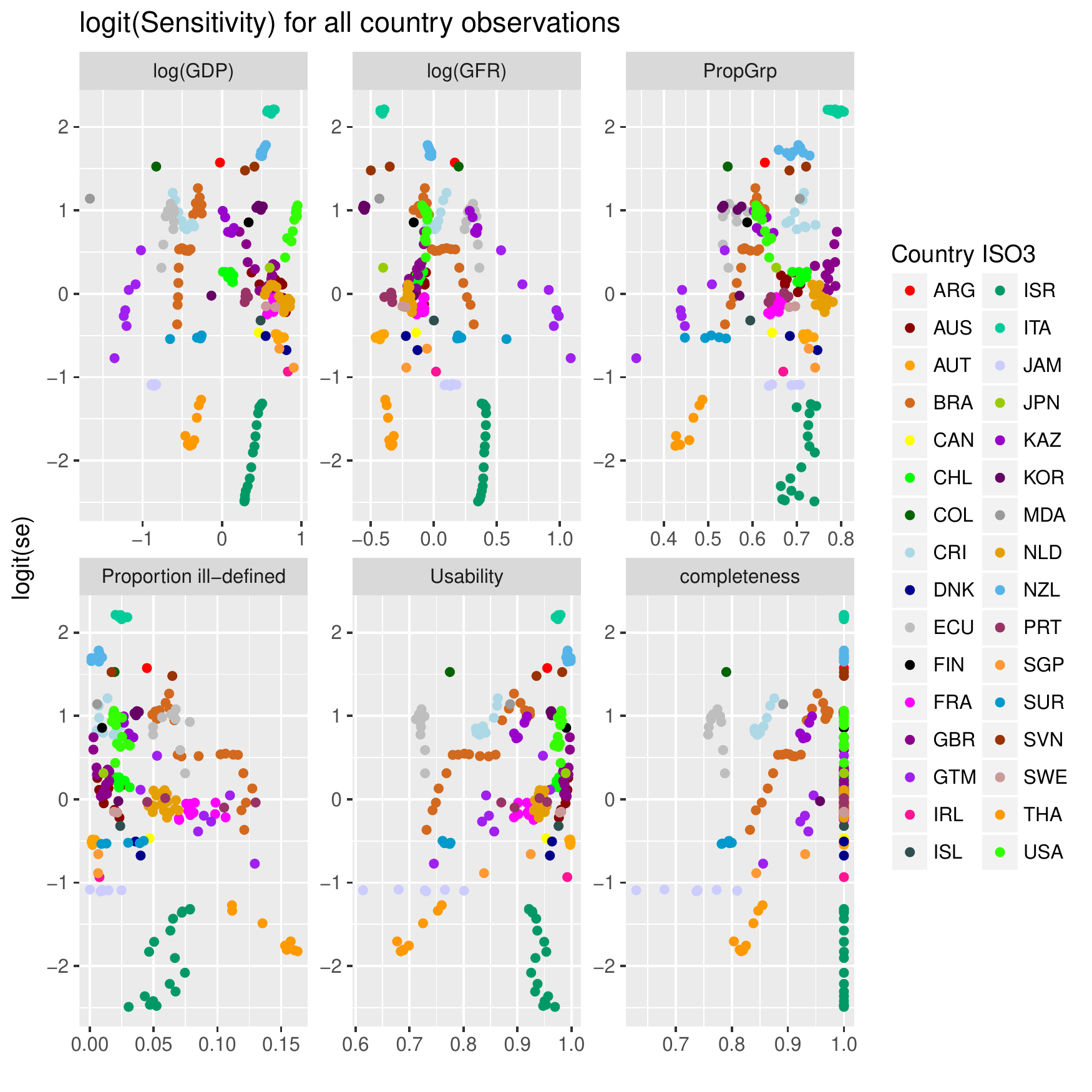}
	\caption{Estimates of sensitivity (on logit-scale) plotted against covariates.}
	\label{fig:cov1}
\end{figure}
\begin{figure}[H]
	\center
	\includegraphics[]{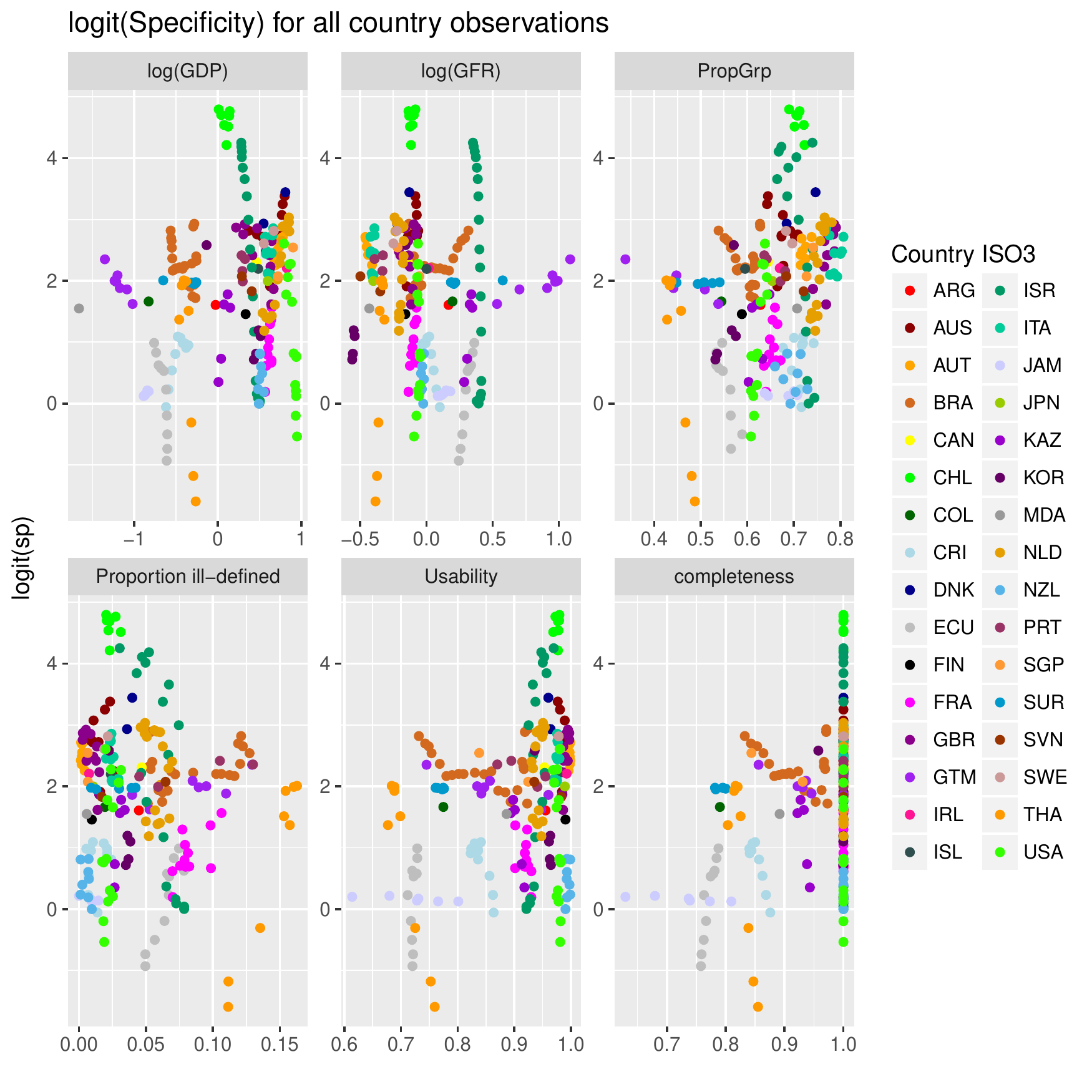}
	\caption{Estimates of specificity (on logit-scale) plotted against covariates.}
	\label{fig:cov2}
\end{figure}
\clearpage

\subsection{BMat 2019}\label{sec:bmat}
\subsubsection{Overview of updates between BMat 2015 and BMat 2019}
UN MMEIG estimates in 2015 were obtained using the UN MMEIG Bayesian maternal mortality estimation model, referred to as BMat (Alkema et al. 2015, Alkema et al. 2017).  In the 2019 estimation round (UN MMEIG 2019), the main change as compared to BMat 2015 was related to the data model for data from CRVS. In addition to this update, we updated the estimation of maternal mortality in crisis years  (UN MMEIG 2019) and updated the data model used for specialized studies data to more accurately incorporate data from smaller studies and account for additional uncertainty in studies that refer to a subset of all deaths to women of reproductive age only (see Section~\ref{sec-spsbmat}).

The approach by which CRVS data are used to inform maternal mortality estimates in BMat 2019 builds upon the model for sensitivity and specificity in CRVS reporting, referred to here as the CRVS model (as explained in the main text in this paper), and BMat 2015 assumptions. In summary, a two-step approach is taken:
\begin{enumerate}
	\item We obtain point estimates of misclassification parameters from the CRVS model, as explained in Section~\ref{sec-crvsforbmat}.
	\item The estimated misclassification parameters are used in BMat for country-years with CRVS data and without specialized studies, see  Section~\ref{sec-bmatdmcrvs}.
\end{enumerate}
We first explain step 2 as it lays out the exact estimates needed from the CRVS model in step 1.

\subsubsection{BMat 2019 data model for CRVS data}\label{sec-bmatdmcrvs}
In BMat 2019, the data model for observed CRVS data is as follows:\\
\begin{eqnarray}
y_{c,t}^{(matCRVS)}|\rho_{c,t}^{(truemat)}, y_{c,t}^{(CRVS)}  &\sim& NegBin\left(E_{c,t}, V_{c,t}\right),\label{eq-bmatlike}
\end{eqnarray}
 where (following notation from the main paper), $y_{c,t}^{(matCRVS)}$ refers to the number of maternal deaths as observed in the CRVS in country $c$ in year $t$, $\rho_{c,t}^{(truemat)}$ is the true probability of a maternal death among all deaths, and $y_{c,t}^{(CRVS)}$ is the total number of deaths registered in the CRVS. \\
\noindent
$E_{c,t}$ and $V_{c,t}$ are defined as follows:
\begin{eqnarray}
E_{c,t} &= & y_{c,t}^{(CRVS)}\cdot\left( \hat{\lambda}^{(+)}_{c,t}\rho_{c,t}^{(truemat)} + \left(1-\hat{\lambda}^{(-)}_{c,t}\right)\left(1-\rho_{c,t}^{(truemat)}\right)\right),\\
V_{c,t} &=& E_{c,t} + y_{c,t}^{2(matCRVS)}\cdot \left(\tilde{V}_{1,c,t} + \tilde{V}_{2,c,t}\right),\\
\tilde{V}_{1,c,t} &=& \hat{v}_{c,t}^{(+)}\cdot \rho_{c,t}^{2(truemat)} + \hat{v}^{(-)}_{c,t}\cdot \left(1-  \rho_{c,t}^{(truemat)}\right)^2 \\
&& - 2\cdot \rho_{c,t}^{(truemat)}\cdot \left(1-\rho_{c,t}^{(truemat)}\right)\hat{u}_{c,t},\\
\tilde{V}_{2,c,t} &=& \hat{m}_{c,t} \cdot \rho_{c,t}^{2(truemat)} \cdot \left(\hat{e}^{(+)}_{c,t} + \hat{e}^{(-)}_{c,t}\right),
\end{eqnarray}
where $\hat{\lambda}^{(+)}_{c,t}$ and $\hat{\lambda}^{(-)}_{c,t}$ refer to point estimates for sensitivity and specificity, $\hat{v}^{(+)}_{c,t}$ and $\hat{v}^{(-)}_{c,t}$ to estimated variances for sensitivity and specificity, $\hat{u}_{c,t}$ to the estimated covariance between sensitivity and specificity, $\hat{e}^{(+)}_{c,t}$ to the estimated squared sensitivity and $\hat{e}^{(-)}_{c,t}$ to estimated squared (1- specificity). Finally, $\hat{m}_{c,t} = 0 $ for country-years with complete CRVS. In country-years with incomplete CRVS, $\hat{m}_{c,t}$ is the estimated variance of $\theta_{c,t}$, with
\begin{eqnarray}
\theta_{c,t} &=& 1/\left(\rho^{(CRVS)}_{c,t} + \left(1-\rho^{(CRVS)}_{c,t}\right)\kappa_{c,t}\right),
\end{eqnarray}
due to uncertainty in the ratio of probabilities of a maternal death among unregistered versus registered deaths $\kappa_{c,t}$ (see Section~\ref{sec:summary}). $\hat{m}_{c,t}$ is approximated using a monte carlo approximation; we set $\hat{m}_{c,t} = Var(\theta_{c,t}^{(h)})$, where samples $\theta_{c,t}^{(h)}$ are constructed as follows:
\begin{eqnarray}
\log\left(\kappa_{c,t}^{(h)}\right) &\sim& N(0, 1),\\
\theta^{(h)}_{c,t} &=& 1/\left(\rho^{(CRVS)}_{c,t} + \left(1-\rho^{(CRVS)}_{c,t}\right)\kappa_{c,t}^{(h)}\right).
\end{eqnarray}
In summary, the variance in $\theta$ is determined by the variability in the ratio of probabilities $\kappa$. The lognormal distribution assigned to $\kappa$ results in first and third quantiles of $\kappa$  around 0.5 and 2, respectively, to reflect the uncertainty associated with this ratio.

The derivation of the data model for CRVS data in Eq.~\ref{eq-bmatlike} is based on the following assumptions:
\begin{eqnarray}
y_{c,t}^{(matCRVS)}|\gamma_{c,t}^{(matCRVS)} &\sim& Poisson\left(\gamma_{c,t}^{(matCRVS)}\cdot y_{c,t}^{(CRVS)}\right),\\
\gamma_{c,t}^{(matCRVS)}|\rho^{(truemat)} &\sim& Gamma(g_1, g_2),
\end{eqnarray}
with $g_1$ and $g_2$ such that $E\left(\gamma_{c,t}^{(matCRVS)}|\rho^{(truemat)}\right) = E_{c,t}/y_{c,t}^{(CRVS)}$ and $V\left(\gamma_{c,t}^{(matCRVS)}|\rho^{(truemat)}\right) = \tilde{V}_{1,c,t} + \tilde{V}_{2,c,t}$.

The data model in Eq.~\ref{eq-bmatlike} specifies which estimates of misclassification parameters are needed to include CRVS-based data into BMat: point estimates $\hat{\lambda}^{(+)}_{c,t}$, $\hat{\lambda}^{(-)}_{c,t}$, (co-)variance estimates $\hat{v}^{(+)}_{c,t}$, $\hat{v}^{(-)}_{c,t}$ and $\hat{u}_{c,t}$, and estimated squared sensitivity $\hat{e}^{(+)}_{c,t}$ and squared (1- specificity) $\hat{e}^{(-)}_{c,t}$. The next sections explain how these estimates were obtained from the CRVS adjustment model, for countries with and without specialized study data.

\subsubsection{Construction of estimates of misclassification parameters for countries with at least one specialized study}\label{sec-crvsforbmat}

The estimates for sensitivity and specificity and associated outcomes need to be informed by all information available regarding misclassification in a country. For the (global) CRVS model as discussed in the main paper, studies were used only if they provided exact information on death counts among deaths that were registered in the CRVS. Studies that reported only on the total number true maternal deaths in country-periods with incomplete CRVS systems, inclusive of missed maternal (U+) deaths, were excluded in the global assessment of misclassification because of lack of information on the relative difference between the true probability of a maternal death among registered versus unregistered deaths. In addition, studies that reported on partial calendar years were excluded. The exclusion decisions were made for the global model to avoid having to make additional assumptions that may affect the global estimates of misclassification. However, for constructing country-specific estimates, we aimed to include all available information, including data points that were excluded from the global model, if inclusion was possible based on reasonable assumptions.

To produce country-specific estimates, using all available data, we obtained country-specific fits of the CRVS model while keeping global parameters fixed at the estimates from the global CRVS model, referred to as a one-country model fit. For each country, all available studies were used, including studies that only provide information that includes missed maternal deaths (explained in Section~\ref{sec-incomplete}), as well as studies that have partial overlap only with CRVS data.
In the one-country model, all model parameters that do not vary across countries or by time are fixed at point estimates from the global CRVS model fit. The process model used for sensitivity and specificity equals the global process model otherwise.

After fitting the one-country model, post-processing of posterior samples of sensitivity and specificity for country-years with CRVS data was carried out. Firstly, for estimates outside observations periods, the variance of sensitivity and specificity was bounded based on global model findings: the maximum variance of sensitivity/specificity  was set to the variance of sensitivity/specificity for a country without specialized study data, as defined in Section~\ref{sec-nostudies}. Secondly, in line with the UN MMEIG 2015 assumptions, we adjusted backprojections of the CRVS-model-based sensitivity and specificity and associated adjustment factors to let sensitivity as well as specificity converge to their associated global value in 5 years. The global point estimates for sensitivity and specificity were set equal to those used for countries without specialized studies, defined in Section~\ref{sec-nostudies} as well.

In summary, estimates of misclassification parameters for countries with at least one specialized study are obtained as follows:
\begin{enumerate}
\item  Fit CRVS model to global data base
\item  Fit CRVS model to all data from country only, using estimates of hyperparameters from the global model fit in step 1.
\item  Post-process posterior samples of sensitivity and specificity from country-specific model fit in step 2, using global estimates of misclassification parameters from global model fit.
\end{enumerate}

\paragraph{Likelihood function for studies counting all maternal deaths in country-periods with incomplete CRVS}\label{sec-incomplete}

\noindent
For a study that reported the total number of true maternal deaths, i.e. those within CRVS plus unregistered maternal deaths, the study-reported count of maternal deaths $z_i^{(truemat)} = z_i^{(T+)} + z_i^{(F-)} + z_i^{(U+)}$ overlaps with CRVS-reported maternal deaths for the corresponding period. Similarly to studies that reported true maternal deaths inside CRVS, refer to Section 4.3, we obtain the exact likelihood function for available death counts by summing over multinomial densities evaluated at each combination $\tilde{\bm{z}}_i = \left(\tilde{z}_i^{(T+)}, \tilde{z}_i^{(T-)}, \tilde{z}_i^{(F+)}, \tilde{z}_i^{(F-)}, \tilde{z}_i^{(U+)}, \tilde{z}_i^{(U-)}\right)$ that satisfied the observed counts. The likelihood function for $i^{th}$ study $(f_i)$ is written as follows:
\begin{align}
f_i = \sum_{\tilde{z}_i^{(U+)}=0}^{z_i^{(UNREG)}} \sum_{\tilde{z}_i^{(T+)}=0}^{z_i^{(matCRVS)}} p_z\left(\tilde{\bm{z}}_{[1:4]}|z_i^{(CRVS)},\bm{\gamma}_{c[i], t[i]}\right) \cdot 1\left(\tilde{z}_i^{(U+)} + \tilde{z}_i^{(T+)} + \tilde{z}_i^{(F-)} = z_{i}^{(truemat)}\right)\cdot k_i \cdot h_i,
\end{align}
\noindent
where $z_i^{(UNREG)}$ refers to the number of unregistered deaths and $p_z\left(\tilde{\bm{z}}_{[1:4]}|z_i^{(CRVS)},\bm{\gamma}_{c[i], t[i]}\right)$ refers to the multinomial density function for the 4 CRVS-based categories from Eq.~\ref{eq-studydm}. To improve computational efficiency and remove combinations that result in values of sensitivity and specificity with negligible probabilities, we added constraints to possible combinations of $\tilde{\bm{z}}_i$, reflected in $k_i$ with
$$k_i = 1\left(\tilde{z}_i^{(T+)} \geq Bin_{2.5\%}(\tilde{z}_i^{(truematCRVS)}, 0.1)\right)\cdot  1\left(\tilde{z}_i^{(T-)} \geq Bin_{2.5\%}(z_{i}^{(CRVS)} - \tilde{z}_i^{(truematCRVS)}, 0.97\right)$$
where $Bin_{2.5\%}(n, p)$ refers to the 2.5th percentile of a Binomial distribution with sample size $n$ and probability $p$, 0.1 is a lower bound for sensitivity, and 0.97 is a lower bound for specificity. Lastly, we included combinations with expected ratios of the proportion maternal inside and outside the CRVS that vary between 0.5 and 2, reflected in $h_i$ with
\begin{eqnarray*}
 h_i &=& 1\left(\tilde{z}_i^{(U+)} \geq Bin_{2.5\%}\left(z_i^{(UNREG)}, p = 0.5 \cdot \frac{\tilde{z}_i^{(truematCRVS)}}{z_i^{(CRVS)}}\right)\right)\\
 && \cdot 1\left(\tilde{z}_i^{(U+)} \leq Bin_{97.5\%}\left(z_i^{(UNREG)},p = 2 \cdot \frac{\tilde{z}_i^{(truematCRVS)}}{z_i^{(CRVS)}}\right)\right).
 \end{eqnarray*}

\subsubsection{Construction of estimates of misclassification parameters for countries without specialized studies}\label{sec-nostudies}

\noindent
We used global estimates of sensitivity, specificity and associated outcomes for all countries without specialized studies, obtained directly from fit of the CRVS model to the global data base, in the BMat data model in Eq~\ref{eq-bmatlike}.  Given the hierarchical set-up of the CRVS model (Eq.~\ref{eq-bhm}), the model can be used directly to produce a predictive distribution of sensitivity and specificity for countries without specialized study data in a reference year. The random walk model (Eqs.~\ref{eq-walk1}) is used for forward and backward extrapolations, and results in constant point estimates of sensitivity and specificity. Specifically, for a country $c^*$ without specialized studies, we set point estimates for sensitivity and specificity equal to their respective global estimates from the global CRVS model fit, $\hat{\lambda}_{c^*,t}^{()} = \hat{\lambda}_{global}^{()}$.  
However, in the bivariate random walk set-up, uncertainty in sensitivity and specificity is increasing as the time lag between the year of interest and the reference year increases, i.e. $Var(\lambda_{c^*,t_{ref}+l}) > Var(\lambda_{c^*,t_{ref}})$ for reference year $t_{ref}$ and time lag $l > 0$. Lacking a natural choice of a reference year for countries without studies, we used constant estimates for the variance, covariance and squared terms, i.e. we set $\hat{v}_{c^*,t}^{()} = \hat{v}_{c^*,t_{ref}+l}^{()}$, $\hat{u}_{c^*,t} = \hat{u}_{c^*,t_{ref}+l}$, and $\hat{e}_{c^*,t}^{()} = \hat{e}_{c^*,t_{ref}+l}^{()}$ for all years $t$, fixed lag $l$ and $t_{ref}$ referring to the year where the hierarchical distribution of Eq.~\ref{eq-bhm} applies. We used a validation exercise to determine the optimal value of time lag $l$, as summarized below, which resulted in the choice of $l = 0$, i.e. to use the uncertainty associated with the distribution of sensitivity and specificity in the reference year (Eq.~\ref{eq-bhm}).

Validation exercise: For each country in the global data set, we predicted its CRVS-based observed PMs using the process model for sensitivity and specificity based on Section~\ref{sec:proc} and hyperparameter estimates from the global model fit, to mimic predicting CRVS-based PMs for a country without specialized study data. For time lags $l = 0, 1, 2, 5, 10, 15$, predictive distributions for each observation year $t$ in the data set were based on the same set of posterior samples of $\lambda_{c^*,t}^{()} = \lambda_{c^*,t_{ref}+l}^{()}$, where $t_{ref}$ refers to the year where the hierarchical distribution of Eq.~\ref{eq-bhm} applies. Per $l$ and per country, we calculated the mean proportion of observations that were outside their respective prediction intervals. Finally, for each choice of $l$, we calculated the mean of the proportions across countries. The results in Table~\ref{tb-valbmat} suggests that even in the reference year, prediction intervals are conservative, with fewer observations falling outside their respective intervals than expected. This finding motivated the use of the reference year for uncertainty estimation. 
\begin{table}[h]
\begin{center}
	\begin{tabular}{|c|c c c c c c  |}

		\hline
		Time lag & 15 & 10 & 5 & 2  & 1 & 0 \\ [0.5ex]
		\hline\hline
		Proportion below 80\% PI & 0.024 & 0.024 & 0.026 & 0.028 & 0.028 & 0.028 \\
		\hline
		Proportion above 80\% PI & 0.009 & 0.009 & 0.027 & 0.030 & 0.04 & 0.05\\
		\hline
	\end{tabular}
	\caption{Summary of validation exercise to determine the choice of time lag $l$ for parameter estimates related to uncertainty in misclassification parameters. PI = prediction interval. }
	\label{tb-valbmat}
\end{center}
\end{table}

\subsubsection{Data model for specialized studies in BMat 2019}\label{sec-spsbmat}

Let specialized studies be indexed by $i$, with the $i$th study referring to country $c[i]$, observation period $t1[i]$ to $t2[i]$ and midpoint $t[i]$. Let  $\rho_{c,t1,t2}^{(truemat)}$ refer to the true probability of a maternal death in country $c$ for the period from $t1$ to $t2$, obtained from the annual probabilities weighted by the total deaths in each year:
$$\rho_{c, t1, t2}^{(truemat)} = \frac{\sum_{t=t1}^{t2}\rho_{c, t}^{(truemat)}y_{c,t}^{(tot)} }{\sum_{t=t1}^{t2}y_{c,t}^{(tot)}}.$$

Data models are discussed separately for studies with complete envelopes $z_i^{(env)} = z_i^{(tot)}$, versus those with incomplete envelopes $z_i^{(env)} < z_i^{(tot)}$.

\paragraph{Studies with complete envelopes}
For specialized study $i$ with envelope $z_i^{(env)} = z_i^{(tot)}$, we assumed
$$z_i^{(truemat)}|\rho_{c[i], t1[i], t2[i]}^{(truemat)} \sim Bin\left(z_i^{(tot)}, \rho_{c[i], t1[i], t2[i]}^{(truemat)}\right),$$
where as before, $z_i^{(truemat)}$ refers to the number of maternal deaths as observed in the specialized study, and $z_i^{(tot)}$ to its respective envelope of all-cause deaths.

\paragraph{Studies with incomplete envelope}
For specialized study $i$ with incomplete envelope $z_i^{(env)} < z_i^{(tot)}$, we assumed (following assumptions and notation from the data model for CRVS data Eq.\ref{eq-bmatlike}):
\begin{eqnarray}
z_i^{(truemat)}|\rho_{c[i], t1[i], t2[i]}^{(truemat)}  &\sim& NegBin(E_i, V_i),
\end{eqnarray}
where, setting $\rho_i = \rho_{c[i], t1[i], t2[i]}^{(truemat)}$ to improve readability,
\begin{eqnarray}
E_i &= & z_i^{(env)}\cdot \left(\hat{\lambda}^{(+)}_{c[i],t[i]}\rho_i+ (1-\hat{\lambda}^{(-)}_{c,t})(1-\rho_i)\right),\\
V_i &=& E_{c,t} +z_i^{2(env)}\cdot (\tilde{V}_{1,i} + \tilde{V}_{2,i}),\\
\tilde{V}_{1,i} &=& \hat{v}_{c[i],t[i]}^{(+)}\cdot \rho_i^2 + \hat{v}^{(-)}_{c[i],t[i]}\cdot (1-  \rho_i^2) \\
&& - 2\cdot \rho_i\cdot (1-\rho_i)\hat{u}_{c[i],t[i]},\\
\tilde{V}_{2,i} &=& \hat{m}_{i} \cdot \rho_i^2 \cdot \left(\hat{e}^{(+)}_{c[i],t[i]} + \hat{e}^{(-)}_{c[i],t[i]}\right),
\end{eqnarray}
where $\hat{m}_{i}$ is the estimated variance of $\theta_{i}$, with
\begin{eqnarray}
\theta_{i} &=& \frac{1}{z_i^{(env)}/z_i^{(tot)}+ \left(1-z_i^{(env)}/z_i^{(tot)}\right)\kappa_{i}},
\end{eqnarray}
due to uncertainty in the ratio of probabilities of a maternal death among uncaptured versus captured deaths $\kappa_{i}$. We set $\hat{m}_{i} = Var(\theta_{i}^{(h)})$, where samples $\theta_{i}^{(h)}$ are constructed as follows:
\begin{eqnarray}
\log\left(\kappa_{i}^{(h)}\right) &\sim& N(0, 1),\\
\theta^{(h)}_{i} &=& \frac{1}{z_i^{(env)}/z_i^{(tot)}+ \left(1-z_i^{(env)}/z_i^{(tot)}\right)\kappa^{(h)}_{i}}.
\end{eqnarray}



\section*{Supplementary material}

**Thailand**
Chandoevwit W, Phatchana P, Sirigomon K, Ieawsuwan K, Thungthong J, Ruangdej S. Improving the measurement of maternal mortality in Thailand using multiple data sources. Popul Health Metr. 2016;14:16. doi 10.1186/s12963-016-0087-z.

**United Kingdom**
Knight M, Kenyon S, Brocklehurst P, Neilson J, Shakespeare J, Kurinczuk JJ (editors) on behalf of MBRRACE-UK. Saving Lives, Improving Mothers' Care: Lessons learned to inform future maternity care from the UK and Ireland Confidential Enquiries into Maternal Deaths and Morbidity 2009–2012. Oxford: National Perinatal Epidemiology Unit, University of Oxford; 2014.

**United Kingdom**
Knight M, Nair M, Tuffnell D, Shakespeare J, Kenyon S, Kurinczuk JJ (editors) on behalf of MBRRACE-UK. Saving Lives, Improving Mothers' Care: Lessons learned to inform maternity care from the UK and Ireland Confidential Enquiries into Maternal Deaths and Morbidity 2013–2015. Oxford: National Perinatal Epidemiology Unit, University of Oxford; 2017.

**United States of America**
CDC's Pregnancy Mortality Surveillance System. From country consultation 2015.

**United States of America**
Creanga AA, Berg CJ, Syverson C, Seed K, Bruce FC, Callaghan WM. Pregnancy-related mortality in the United States, 2006-2010. Obstet Gynecol. 2015;125(1):5-12.

**Uruguay**
Data received for 2000-2017 during the country consultation 2019.

End of Box 1.

**Box 2: SEARCH TERMS**

EMBASEw
http://www.embase.com
No AGE, HUMAN
YEAR limits applied : [1990-2050]/py
Options Also search as free text was enabled.

| # | Searches | Results |
|---|---|---|
| 1 | 'maternal mortality'/exp OR 'maternal mortality' OR 'maternal mortalities' | 22,873 |
| 2 | 'underreporting' OR 'under reporting' OR underreported OR 'under reported' OR 'data quality' OR 'official figures' OR 'record linkage' OR 'quality of information' OR 'officially reported' OR 'multiple sources' OR 'linkage' OR 'under registered' OR 'under registration' OR underregistered OR underregistration OR 'under registering' OR 'source of error' OR 'misclassification' OR 'misclassified' OR (errors AND ('registration'/exp OR registration)) OR 'late maternal mortality' OR 'confidential enquiries' OR 'confidential enquiry' | 180888 |
| 3 | 'data collection method'/exp OR 'health survey'/exp AND (standard* OR method*) | 558644 |
| 4 | #1 AND (#2 OR #3) | 1760 |
| 5 | 'pregnancy'/exp OR 'pregnancy complication'/exp OR 'pregnancy disorder'/exp OR 'abortion'/exp AND ('death'/exp OR deaths OR 'mortality'/exp OR fatal OR fatalities OR deceased) | 105286 |
| 6 | #2 AND #5 | 1143 |
| 7 | #4 OR #6 | 2440 |
| 8 | #4 OR #6 AND [1990-2016]/py | 2335 |

PubMed
http://www.pubmed.gov
Filters: Publication date from 1990/01/01 to 2050/12/31

| # | Searches | Results |
|---|---|---|
| 1 | "underreporting"[tiab] OR "under reporting "[tiab] OR underreported [tiab]OR " under reported" [tiab]OR "data quality" [tiab] OR "official figures" [tiab] OR "record linkage" [tiab] OR "quality of information" [tiab] OR "officially reported "[tiab] OR "multiple sources" [tiab] OR" linkage" [tiab] OR "under registered" [tiab] OR" under registration" [tiab] OR "under registering" [tiab] | 209385 |

| | | |
|---|---|---|
| | OR underregistered[tiab] OR underregistration[tiab] OR "under registering" [tiab] OR "source of error" [tiab] OR "misclassification" [tiab] OR "misclassified "[tiab] OR (errors[tiab] AND registration[tiab]) OR "late maternal mortality" [tiab] OR "confidential enquiries" [tiab] OR "confidential enquiry" [tiab] OR "Data Collection/methods"[Mesh] OR "Data Collection/standards"[Mesh] OR "Population Surveillance/methods"[Mesh] OR "Population Surveillance/standards"[Mesh] | |
| 2 | "Maternal Mortality"[Mesh] OR "maternal mortality" [Tw] OR "maternal mortalities" [Tw] OR ((Pregnancy[mesh] OR "pregnancy complications" [Mesh] or "pregnant women" or parturition[mesh] or mothers[mesh] or "maternal health services"[mesh] or pregnancy or pregnant or parturition or mother* or gestation or gestational or childbirth or childbirths or maternal or maternity ) AND (mortality OR mortalities OR Death OR deceased OR fatality OR fatalities)) | 116919 |
| 3 | #1 AND #2 | 1977 |
| 4 | "mothers"[MeSH Terms] OR "mothers"[All Fields] OR "mother"[All Fields] OR "mothers"[MeSH Terms] OR "mothers"[All Fields] OR "maternal"[All Fields] OR "pregnancy"[MeSH Terms] OR "pregnancy"[All Fields] OR "parturition"[MeSH Terms] OR "parturition"[All Fields] OR "postpartum period"[MeSH Terms] OR "postpartum"[All Fields] AND "period"[All Fields] OR "postpartum period"[All Fields] OR "postpartum"[All Fields] OR antepartum[All Fields] OR intrapartum[All Fields] OR "parturition"[MeSH Terms] OR "parturition"[All Fields] OR "childbirth"[All Fields] OR "delivery, obstetric"[MeSH Terms] OR ("delivery"[All Fields] AND "obstetric"[All Fields]) OR "obstetric delivery"[All Fields] OR "parturition"[MeSH Terms] OR "parturition"[All Fields] OR "birth"[All Fields] OR termination[All Fields] OR "abortion, induced"[MeSH Terms] OR ( "abortion"[All Fields] AND "induced"[All Fields]) OR "induced abortion"[All Fields] OR "abortion"[All Fields] OR "abortion, spontaneous"[MeSH Terms] OR ("abortion"[All Fields] AND "spontaneous"[All Fields]) OR "spontaneous abortion"[All Fields] OR "miscarriage"[All Fields] | 566626 |
| 5 | "death"[MeSH Terms] OR "death"[All Fields] OR fatal[All Fields] OR fatality[All Fields] OR "mortality"[Subheading] OR "mortality"[All Fields] OR "mortality"[MeSH Terms] | 1543760 |
| 6 | #4 AND #5 AND #1 | 1715 |
| 7 | #6 OR #3 | 2516 |
| 8 | #6 OR #3 Publication date from 1990/01/01 to 2050/12/31 | 2285 |

Global Index Medicus
http://www.globalhealthlibrary.net
No AGE, HUMAN or YEAR limits applied.
Options : Regional Indexes searched

| # | Searches (LILACS) | Results |
|---|---|---|
| 1 | ((underreporting) OR (under reporting ) OR underreported OR ( under reported) OR (data quality) OR (official figures) OR (record linkage) OR (quality of information) OR (officially reported ) OR (multiple sources) OR ( linkage) OR (under registered) OR ( under registration) OR (under registering) OR underregistered OR underregistration OR (under registering) OR (source of error) OR (misclassification) OR (misclassified ) OR (errors AND registration) OR (late maternal mortality) OR (confidential enquiries) OR (confidential enquiry)) AND ((MOTHERS AND Mortality) OR (Maternal Mortality) OR (Maternal Death) OR (maternal mortality) OR (maternal deaths) OR (pregnancy related deaths) OR (pregnancy related deaths)) | 910 |
| | IMEMR (same as above) | 163 |
| | WPRIM (same as above) | 33 |
| | IMSEAR(same as above) | 20 |
| | AIM(same as above) | 16 |

EBSCO
http://www.ebsco.com
No AGE, HUMAN or
YEAR limits applied 1990 – 2013
Searched in SUBJECTS, ABSTRACT and TITLE fields only across databases suite.

| # | Searches | Results |
|---|---|---|
| 1 | (underreporting OR under reporting OR underreported OR under reported OR data quality OR official figures OR official national figures OR record linkage OR quality of information OR officially reported OR multiple sources OR linkage OR under registered OR under registration OR under registering OR underregistered OR underregistration OR under registering OR sources of error OR misclassification OR misclassified OR (errors AND registration) OR late maternal mortality OR confidential enquiries OR confidential enquiry OR (data collection AND (methods OR standards)) OR audit OR (population surveillance AND (methods OR standards))) AND (maternal mortality OR pregnancy related deaths OR maternal deaths) | 2831 citations found. EBSCO system Duplicates removed |

| | | remaining : 1019<br><br>See below for results per database |
|---|---|---|
| | Academic Search Complete | 550 |
| | Academic Search Premier | 535 |
| | CINAHL Complete | 457 |
| | CINAHL Plus with Full Text | 428 |
| | Health Source: Nursing/Academic Edition | 121 |
| | Women's Studies International | 117 |
| | Gender Studies Database | 93 |
| | Consumer Health Complete - EBSCOhost | 81 |
| | PsycINFO | 80 |
| | Food Science Source | 61 |
| | SocINDEX with Full Text | 31 |
| | MasterFILE Premier | 30 |
| | Business Source Complete | 29 |
| | Public Affairs Index | 27 |
| | Business Source Premier | 26 |
| | Environment Complete | 18 |
| | Psychology and Behavioral Sciences Collection | 16 |
| | Vocational and Career Collection | 12 |
| | MedicLatina | 11 |
| | Middle Eastern & Central Asian Studies | 11 |
| | Health Source - Consumer Edition | 9 |
| | Education Research Complete | 9 |
| | Agricola | 8 |

| | | |
|---|---|---|
| | [Peace Research Abstracts](#) | 8 |
| | [Alt HealthWatch](#) | 7 |
| | [Professional Development Collection](#) | 6 |
| | [Education Full Text (H.W. Wilson)](#) | 6 |
| | [International Security & Counter Terrorism Reference Center](#) | 5 |
| | [SPORTDiscus with Full Text](#) | 5 |
| | [Risk Management Reference Center](#) | 3 |
| | [Historical Abstracts](#) | 3 |
| | [Political Science Complete](#) | 3 |
| | [ERIC](#) | 2 |
| | [Computer Source](#) | 2 |
| | [Communication & Mass Media Complete](#) | 2 |
| | [Library, Information Science & Technology Abstracts with Full Text](#) | 2 |
| | [Computers & Applied Sciences Complete](#) | 2 |
| | [Associates Programs Source](#) | 2 |
| | [Vocational Studies Premier](#) | 2 |
| | [Caribbean Search](#) | 2 |
| | [Criminal Justice Abstracts with Full Text](#) | 2 |
| | [Biological & Agricultural Index Plus (H.W. Wilson)](#) | 2 |
| | [Legal Collection](#) | 1 |
| | [Bibliography of Native North Americans](#) | 1 |
| | [National Criminal Justice Reference Service Abstracts](#) | 1 |
| | [Central & Eastern European Academic Source](#) | 1 |
| | [Humanities Abstracts (H.W. Wilson)](#) | 1 |
| | [Humanities Full Text (H.W. Wilson)](#) | 1 |
| | Total of citations found after duplicates removed | 1019 |

Web of Science
http://www.webofknowledge.com
No AGE, HUMAN or YEAR limits applied.

| # | Searches (TOPIC FIELD) | Results |
|---|---|---|
| 1 | (underreporting OR "under reporting" OR underreported OR "under reported" OR "data quality" OR "official figures" OR "official national figures" OR "record linkage" OR "quality of information" OR "officially reported " OR "multiple sources" OR linkage OR "under registered" OR "under registration" OR "under registering" OR underregistered OR underregistration OR "under registering" OR "sources of error" OR misclassification OR misclassified OR (errors AND registration) OR "late maternal mortality" OR "confidential enquiries" OR "confidential enquiry" OR ("data collection" AND (methods OR standards)) OR audit OR ("population surveillance" AND (methods OR standards))) AND ("maternal mortality" OR "pregnancy related deaths" OR "maternal deaths") | 688 |

Popline
http://www.popline.org
.

| # | Searches (TOPIC FIELD) | Results |
|---|---|---|
| 1 | (underreporting OR "under reporting" OR underreported OR "under reported" OR "data quality" OR "official figures" OR "official national figures" OR "record linkage" OR "quality of information" OR "officially reported " OR "multiple sources" OR linkage OR "under registered" OR "under registration" OR "under registering" OR underregistered OR underregistration OR "under registering" OR "sources of error" OR misclassification OR misclassified OR (errors AND registration) OR "late maternal mortality" OR "confidential enquiries" OR "confidential enquiry" OR ("data collection" AND (methods OR standards)) OR audit OR ("population surveillance" AND (methods OR standards))) AND ("maternal mortality" OR "pregnancy related deaths" OR "maternal deaths") | 142 |

Web of Science
http://www.webofknowledge.com
No AGE, HUMAN or YEAR limits applied.

| # | Searches (Web of Science – Russian Index) | Results |
|---|---|---|
| 1 | («недостаток информации» OR «утерянные данные» OR «дефекты сбора данных» "несообщение" OR «сокрытие» OR «несообщённый» OR «сокрытый» OR «несообщённая» OR «сокрытая» OR «скрытый» OR «скрытая» OR «сокрытые» OR «скрытые» OR «качество информации» OR | |

| | | |
|---|---|---|
| | «качество данных» OR «официальные данные» OR «официальные цифры» OR «официальная статистика» OR «национальная статистика» OR «национальные данные» OR «многочисленные источники» OR «множественные источники» OR «сцепленные данные» OR «связанные данные» OR «незарегистрированные» OR «незарегистрированный» OR «незарегистрированная» OR «не зарегистрированный» OR «не зарегистрированная» OR «не зарегистрированные» OR «отказ от регистрации» OR «регистрация не проводилась» OR «не регистрировалась» OR «не регистрировался» OR «не регистрировались» OR «причина ошибки» OR «источник ошибки» OR «причины ошибки» OR «источники ошибки» OR «причина ошибок» OR «источник ошибок» OR «причины ошибок» OR «источники ошибок» OR «ошибочная классификация» OR «ошибка классификации» OR «ошибка в классификации» OR «неправильная классификация» OR «неверная классификация» OR «неправильная группировка» OR «неверная группировка» OR «ошибка в группировке» OR «ошибочная группировка»  OR «поздняя материнская смертность» OR «поздней материнской смертности» OR «позднюю материнскую смертность» OR «конфиденциальный запрос» OR «конфиденциальное расследование» OR «закрытая информация» OR «закрытые сведения» OR «закрыть информацию» OR «утаить информацию» OR «утаённая информация» OR («сбор информации» OR «сбора информации» AND («методы» OR «стандарты» OR «механизм» OR «техника» OR «алгоритм» OR «методика») OR «аудит» OR («надзор» OR «популяционный надзор» OR «здоровье населения» OR «состояние здоровья населения» OR «здоровье популяции»)) | |
| 2 | («материнская смертность» OR «акушерская смертность» OR «акушерско-гинекологическая смертность» OR «послеродовая смертность» OR «смерть в родах» OR «родовая смертность» OR «гибель рожениц» OR «гибель родильниц» OR «смертность рожениц» OR «смертность родильниц») | |
| | 1 AND 2 | 18 |

**End of Box 2**



\end{document}